\documentclass[dvips,10pt,cmp,aps,twocolumn,nofootinbib]{revtex4}

\usepackage{amsmath,amssymb,epsf,epsfig,amsthm,bm,graphicx,subfigure}

\newcommand{\cL}{{\cal L}}

\newcommand{\cK}{{\cal K}}

\newcommand{\bo}{\bar{\omega}}
\newcommand{\bE}{\bar{E}}
\newcommand{\bD}{\bar{D}}
\newcommand{\bO}{\bar{\Omega}}

\newcommand{\bT}{\bar{T}}

\theoremstyle{theorem}
\newtheorem{theorem}{Theorem}[section]

\newtheorem{corollary}[theorem]{Corollary}

\theoremstyle{definition}

\newtheorem{definition}[theorem]{Definition}

\newtheorem{question}[theorem]{Question}
\newtheorem{convention}[theorem]{Convention}

\begin{document}

\title{Singular sources in gravity and homotopy in the space of connections}%

\author{E. Gravanis\\ }
\affiliation{}
\author{S. Willison}
\email{steve@cecs.cl} \affiliation{Centro de Estudios Cient\'ificos
(CECS), Casilla 1469, Valdivia, Chile}
\date{8 January 2009}

\begin{abstract}
Suppose a Lagrangian is constructed from its fields and their
derivatives. When the field configuration is a distribution, it is
unambiguously defined as the limit of a sequence of smooth fields.
The Lagrangian may or may not be a distribution, depending on
whether there is some undefined product of distributions. Supposing
that the Lagrangian is a distribution, it is unambiguously defined
as the limit of a sequence of Lagrangians. But there still remains
the question: Is the distributional Lagrangian uniquely defined by
the limiting process for the fields themselves? In this paper a
general geometrical construction is advanced to address this
question. We describe certain types of singularities, not by
distribution valued tensors, but by showing that the action
functional for the singular fields is (formally) equivalent to
another action built out of \emph{smooth} fields. Thus we manage to
make the problem of the lack of a derivative disappear from a system
which gives differential equations. Certain ideas from homotopy and
homology theory turn out to be of central importance in analyzing
the problem and clarifying finer aspects of it.

The method is applied to general relativity in first order
formalism, which gives some interesting insights into distributional
geometries in that theory. Then more general gravitational
Lagrangians in first order formalism are considered such as Lovelock
terms (for which the action principle admits space-times more
singular than other higher curvature theories).
\\

Preprint: CECS-PHY-09/01

\end{abstract}

\maketitle


\section{Introduction}

There are many situations in physics where a singular or non-smooth
field is introduced as an approximation or limiting case of some
smooth field. An elementary example is an electrically charged plate
with surface charge density $\sigma_e$. The Maxwell equations tell
us that the divergence of the electric field is equal to the
electric charge density $\rho_e$. Letting $z$ be a coordinate
orthogonal to the plate pointing from left to right, we can
integrate across the infinitesimal width of the plate, say between
$z = -\epsilon$ and $+\epsilon$ to get:
$\int_{-\epsilon}^{+\epsilon} dz \left(\partial_z E^z +
\cdots\right)
 = \int_{-\epsilon}^{+\epsilon} dz \rho_e\ \Rightarrow $
\begin{equation}
 \quad \left[\left[E^z \right]\right] = \sigma_e\, ,
\end{equation}
where the last statement follows from taking the limit $\epsilon \to
0$. In this limit the charge per unit area $\sigma_e :=
\lim_{\epsilon \to 0} \int_{-\epsilon}^{+\epsilon} \rho_e$ is the
integral of a singular charge density. Square brackets are used to
denote the jump in a quantity: $\left[\left[E^z\right]\right] :=
(E^z)_R -(E^z)_L$ is the normal component of the electric field
evaluated on the right hand side of the plate minus the same
quantity on the left. Here the charge density is a Dirac delta
distribution and the electric field is a Heaviside distribution. The
field equations then are well defined, using the mathematical theory
of distributions, even in the idealised case when the plate has zero
thickness. Finally we note that, when considering the energy density
or the Lagrangian, one comes across $(E^z)^2$ which is somewhat
ambiguous in the thin shell limit, so it is necessary to know
something about the internal arrangement of charge inside of the
shell\footnote{We need to know that there is not some wild
oscillation between large positive and negative values of $E^z$
inside the shell which integrates to zero but whose square does not.
Then we can legitimately replace Heaviside function squared with the
uniform function 1.} in order to determine these quantities.

The analogous problem of a massive shell in general relativity,
although considerably more subtle, is also well known. Suppose that
we have a shell, whose world-volume $\Sigma_{12}$ is a non-null
hypersurface with some singular stress tensor living on it. The
metric is assumed to be continuous across the shell in some
appropriate coordinate system but not necessarily differentiable.
Let us parametrise the location of $\Sigma_{12}$ by $\Phi(x^\mu)= 0$
and introduce coordinates $\xi^i$ intrinsic to the shell. The
Lanczos equation for a singular shell is, in covariant
form~\cite{Israel-66}:
\begin{equation}
 \left[[ K_{ij} - h_{ij}K\right]] = - 8\pi G_N  S_{ij}\, .
\end{equation}
where $K_{ij}$ is the extrinsic curvature tensor, $h_{ij}$ is the
intrinsic metric of the shell world-volume and as before $[[ \cdots
]]$ denotes the jump. This is obtained by integration of the
Einstein equations $G_{\mu\nu} = 8 \pi G_N T_{\mu\nu}$ across the
shell, assuming a stress tensor which is a Dirac delta distribution
$T_{\mu\nu} = e_\mu^i e_\nu^j S_{ij} \delta(\Phi)$. Now it happens
that the Lanczos equations can also be obtained from an action
principle (this point was emphasised in Ref.~\cite{Hayward-90}). Let
us assume that the shell has no boundary and divides the space-time
into two bulk regions $\Sigma_1$ and $\Sigma_2$. The gravitational
action is the sum of the York action $ \int_{\Sigma} R \sqrt{-g}\,
d^4 x \mp 2 \int_{\partial \Sigma}\! K \sqrt{\mp h}\, d^3 x$ for
each bulk region. The two surface terms combine to give the jump in
the trace of the extrinsic curvature across the shell:
 \begin{equation}\label{JumpK} \mp 2
\int_{\Sigma_{12}} (K_2 -K_1) \sqrt{\mp h}\, d^3 x
\end{equation}

We observe that if there is no matter $S_{ij} =0$, the term
(\ref{JumpK}) simply imposes a condition on the differentiability of
the metric: that there exists a coordinate system where the metric
is at least once differentiable in the direction normal to the
shell.
The surface term (\ref{JumpK}) can be inserted for free into the
gravitational action on any non-null hypersurface, something which
is important in the path integral formulation of quantum gravity.
The insertion of (\ref{JumpK}) on a spacelike hypersurface
corresponds to inserting a complete set of states~\cite{Hawking-79}.

So then the Lanczos equation, derived by integration of the
distributional field equation, can also be obtained from an action
with a surface term which, unsurprisingly, is the integral of the
distributional Lagrangian~\cite{Hayward-90}. In this paper we wish
to focus mainly on Lagrangians which are distributions or singular
in some suitably mild way. In some theories of gravity, the
possibility of distributional fields enters at the level of the
classical action principle. And certainly, in the path integral
formulation they are expected to be important~\cite{Isham-75}.

The Lagrangian is a function of fields which are themselves not
smooth functions. In the case of general relativity, the fields are
the components of the metric, which is continuous but not
necessarily differentiable, and its `derivatives', which are
actually distributions. Let us write generally ${\cal L}= {\cal
L}(\Psi)$, where $\Psi$ is a field or collection of fields. If
$\Psi$ is a distribution it can always be represented as a limit
$\lim_{\alpha \to \infty}\Psi_{\alpha}$ of a family of smooth fields
$\{\Psi_\alpha\}$ (e.g. a Heaviside distribution can be realised as
the limit $\alpha \to \infty$ of functions $\Psi_{\alpha}(z) = 2
\arctan (z \alpha)/\pi$). The question is:

\begin{question}
 Under what conditions is
 \begin{equation}\label{unambiguous}
    {\cal L} = \lim_{\alpha \to \infty} {\cal L}(\Psi_\alpha)
 \end{equation}
 unambiguously defined? By unambiguously defined we mean that for
 \emph{any} family
 $\Psi_1, \Psi_2, \dots,\Psi_\alpha,\dots$ which converges to $\Psi$, ${\cal
 L}(\Psi_\alpha)$ always converges to ${\cal L}$ as $\alpha \to \infty$.
\end{question}

Although we will mainly focus on the Lagrangian, one can ask the
same question about the field equations:
\begin{question}
 Under what conditions is
 \begin{equation}\label{unambiguous_Field_Equation}
     \lim_{\alpha \to \infty} \frac{\delta}{\delta \Psi_\alpha}
    {\cal L}(\Psi_\alpha)
 \end{equation}
 unambiguously defined?
\end{question}

Above, the Lagrangian and field equations are to be understood as
distributions or generalised functions, well defined under
integration in a way that will be made more precise later. We have
chosen to define ${\cal L}$ by first forming the function and then
taking the limit. That is, we use the limiting process for $\Psi$ to
\emph{replace} ${\cal L}(\Psi)$ by a distribution. Instead of
question 1.1. we could have asked: `Under what conditions is ${\cal
L}(\Psi)$ defined as the function of non-smooth fields?' This is not
quite the same thing since generally ${\cal L}(\lim_{\alpha\to
\infty} \Psi_\alpha)$ is not the same as (\ref{unambiguous}) and may
not exist even when (\ref{unambiguous}) is well defined.

There are good reasons for defining the Lagrangian and field
equations the way we have: Firstly, if $\Psi$ is an approximation to
a smooth field, as would be the case for a thin shell, then the
unambiguity of (\ref{unambiguous}) and
(\ref{unambiguous_Field_Equation}) means that, for a sufficiently
thin shell, the details of the internal structure become irrelevant
to the physical description (at scales much larger than the
thickness of the shell) and so the distributional field captures all
the relevant physics to a good approximation.

Secondly, suppose one wants to admit non-smooth $\Psi$ as exact
fields entering in the classical variational principle. The
important thing is to find a mathematically well defined way to do
so, which may or may not be through the theory of distributions
(defined through linear functionals on the space of smooth test
functions). There are various approaches to defining generalised
functions in generally covariant theories (see
Ref.~\cite{Steinbauer-06} for a review). The important thing is to
find an unambiguous prescription and our approach seems natural in
view of the thin shell limit process described above.

Thirdly, if we suppose that the path integral for gravity is truly
some kind of sum over classical metrics, then it is inevitable that
these kind of limits occur. Let us look at the restriction of the
sum to a family of smooth fields: $\sum_\alpha  \exp(i/\hbar \int
{\cal L}(\Psi_\alpha))$. The `last term' in the series is
$\exp(i/\hbar \int {\cal L})$ with ${\cal L}$ defined as in
(\ref{unambiguous}). The question arises whether the non-smooth
$\Psi$ is to be regarded as a single point in the space of fields.
That is, if there is some other sequence of fields $\Psi'_1,
\Psi'_2, \dots, \Psi'_\alpha,\dots$, which converges to the same
distribution $\Psi$, does the limiting term contribute the same
value to the path integral? If not, we can not be justified in
identifying the two limits as the same point in the space of fields.
This is question 1.1. Furthermore, is the action slowly varying in
the vicinity of the limit point? If so then the method of stationary
phase may be applicable. This is related to question 1.2. Although
the status of the path integral for gravity is highly questionable,
these considerations lends some support for our definitions of the
classical variational problem with non-smooth fields.

It is apparent that the answer to questions 1.1 and/or 1.2. will be:
`not for every type of non-smooth field'. A very similar question to
1.2 was addressed by Geroch and Traschen in the case of general
relativity \cite{Geroch-87}. They analysed under what conditions the
Riemann tensor (and contractions of it) was an unambiguous
distibution in terms of the limit of a family of smooth metrics.
They found an interesting example where the Einstein equations are
ambiguous: the straight singular cosmic string. There are various
limiting processes for the metric which yield the same string
metric, but when Einstein's equations are considered, give different
expressions for the mass per unit length of the string. For a shell
of codimension one, the limiting process is unambiguous and always
yields the Lanczos equations given
above~\cite{Israel-66}\cite{Geroch-87}.

In higher dimensions, one can generalise to the Lovelock
gravitational action~\cite{Lovelock-71}. In that case it is also
found that the limiting process for a thin shell leads to
unambiguous Lagrangian and field
equations~\cite{Fursaev-95}\cite{Deruelle-03}\cite{Gravanis-07}.
Likewise for the intersection of shells, without deficit
angle~\cite{Gravanis-03}\cite{Gravanis-04}. It has been shown that
in some cases these results generalise to the first order
formulation with non-vanishing
torsion~\cite{Giacomini-06}\cite{Willison-04}.

In this paper we shall generalise greatly. The method we introduced
in Refs.~\cite{Gravanis-03}\cite{Gravanis-04} and shall further
develop here relies on concepts more commonly used in gauge theory
rather than in gravity. The method can be applied a to very general
class of theories, although useful results are expected for theories
constructed along the lines of Ref.~\cite{Regge-86} from
differential forms and their exterior derivatives, without Hodge
dual. Examples shall focus on various different theories of gravity
in first order formalism with or without torsion~\cite{Mardones-91}.

In the next section we introduce the necessary ideas. Then in we
shall proceed to define rigorously the mathematical machinery
needed. In section III we shall consider some specific applications
to gravity theories (with torsion). Then in sections IV and V we
shall develop further the mathematical formalism. Section VI
contains some concluding remarks.

\section{Geometrical Construction}

We are dealing really with the topology of the space of all fields,
 be it the space of semi-Riemannian metrics for
gravity or the space of gauge connections for gauge theory, etc. In
gauge theory there are some elegant and powerful results in this
direction. To make use of these, we shall formulate the problem in a
gauge-theoretic way. The gravitational action will be regarded as a
functional of the spin connection $\omega^{\hat{a}\hat{b}}_{\ \
\mu}$ and vielbein $E^{\hat{a}}_{ \mu}$, which are one-forms and
transform under the local Lorentz transformations as a connection
and vector respectively. It is convenient to drop the local Lorentz
and space-time indices and just write $\omega$, $E$.

The example of the Lanczos equation shows what we would like to do.
The limit of a sequence of fields is replaced by another kind of
limit: a directional limit. The spacetime is piecewise smooth and we
have a single connection $\lim_{\alpha \to \infty} \omega_\alpha$
which is undefined on $\Sigma_{12}$ and it is effectively replaced
by two connections on $\Sigma_{12}$, given by the directional
limits: let $p$ be a point on $\Sigma_{12}$
\begin{equation*}
 \omega_1 (p) := \lim_{x \in \Sigma_1 \to p} \omega(x)\, ,
 \qquad \omega_2(p) :=
 \lim_{x \in \Sigma_2 \to p} \omega(x)\, .
\end{equation*}
The above makes sense provided that $\omega$ in each bulk region can
be extended smoothly in some neighbourhood across the shell. When
considering a piece-wise smooth space-time, effectively we have not
one smooth connection in this neighbourhood, but rather two.

More generally, there may be a network of singular hypersurfaces
dividing up the spacetime into many smooth bulk regions, labeled
$\Sigma_i$, $i =1 , \dots, N$. In the interior of each bulk region
space-time is smooth. In the neighbourhood of an intersection of
hypersurfaces we have a whole collection of smooth vielbeins and
spin connections $ \{\omega_i, E_i\}$ associated with all of the
$\Sigma_i$ meeting at that intersection. [It is important to
remember that here and in what follows the index $i$ labels bulk
regions and is not to be mistaken for a tensor index.]

Now there is a novel geometrical way to include such a collection of
fields in an action principle. To do this we first need to introduce
some general results about the topology of the space of connections
in gauge theory, especially those related to secondary
characteristic classes. Then we will see how this method is applied
to the vielbein and other fields.

\subsection{The Idea}\label{Idea_section}

In gauge theory, the basic field is a gauge field $A$, transforming
as $A \to A^g := g^{-1}(A + d)g$ under a gauge transformation, where
$g(x) \in G$ is an element of the gauge group. The phase space is
${\mathbb A}$, the space of connections (or, to be precise the
physical phase space is ${\mathbb A}/{\cal G}$ where ${\cal G}$ is
the group of gauge transformations).

Now let $A_1, A_2 \dots , A_N$ be a set of gauge fields. Then the
linear combination $A_t := A_1 t + A_2 (1-t)$ is also a connection:
under the simultaneous gauge transformations $A_1 \to A_1^g$, $A_2
\to A_2^g$, one can verify that $A_t$ transforms in the correct way
$A_t \to A_t^g = g^{-1}(A_t +d)g$. More generally, the combination
\begin{equation*}
 {\bar A} := t^1 A_1 + \cdots + t^N A_N,  \qquad
 (t^1 + \cdots + t^N =1)
\end{equation*}
is a connection. It is simple to check this but it tells us
something quite important about the space of connections. This
construction explicitly demonstrates that ${\mathbb A}$ is
contractible (see Ref. \cite{Nash-91} for a detailed introduction to
the topology of $\mathbb{ A}$).

It is natural to restrict the parameters $t^i$ to $0\leq t^i \leq 1\
\forall i$, so that they form the coordinates of a convex simplex
$S$ in $\mathbb{A}$. Also, we introduce the exterior derivative
$\delta := dt^\alpha \frac{\partial}{\partial t^\alpha }$ on $S$. We
can introduce the curvatures $\bar{ F} := d\bar{A} + \bar{A} \wedge
\bar{A}$ and ${\cal F} := (d+\delta )\bar{A} + \bar{A} \wedge
\bar{A} = \delta \bar{A}+\bar{F}$. These curvatures are useful in
the construction of secondary characteristic
classes~\cite{Chern-74}\cite{Chern-Book}. The secondary
characteristic form is:\footnote{For convenience we drop the wedge
notation and use ${\cal F}^n$ as shorthand for the $n$-fold wedge
product ${\cal F} \wedge \cdots \wedge {\cal F}$.}
\begin{equation*}
 \langle {\cal F}^n\rangle
 =\langle (\delta {\bar A}+{\bar F})^{n}\rangle
  = \langle \bar{F}^n + n\, \delta {\bar A}\,{\bar F}^{n-1}
 + \cdots + (\delta {\bar A})^n
  \rangle \, ,
\end{equation*}
where $\langle \cdots \rangle$ denotes an invariant `Trace'. Then it
is easy to prove that $(\delta +d)\langle {\cal F}^n\rangle =0$.
This has found various applications in physics, for example the
descent equations which describe non-abelian
anomalies\cite{Guo-84}\cite{Manes-85}\cite{Alvarez-85}.
\\

What has all this got to do with the problem of well-defined gravity
actions for distributional geometry? Let space-time be defined on a
manifold $M$ of dimension $D$. We shall take as our gravity
Lagrangian a polynomial in curvature two-form, torsion two-form and
vielbein one-form. We shall denote this as ${\cal L }(E,\omega) =
\langle \Omega^p T^q E^s \rangle$. The angled bracket means that we
must contract with a Lorentz invariant tensor. There are two
options:
\\
i) We contract with the totally antisymmetric Levi-Civita tensor. In
this case $q$ must be zero and we have a Lagrangian of Lovelock
gravity;
\\
ii) We contract with some combination of Minkowski metrics. This
leads to the more exotic types of action considered in Ref.
\cite{Mardones-91}. \\Note that, since ${\cal L}$ is a $D$-form,
there is the constraint $2p + 2q + s = D$.

The action for a smooth geometry is
\begin{gather}
 I[E, \omega] = \int_M {\cal L}(E,\omega).
\end{gather}

When considering a piece-wise smooth space-time, effectively we have
not one vielbein and one connection, but rather a whole collection
of smooth fields $\{E_i, \omega_i\}$ associated with each bulk
region $\Sigma_i$. Inspired by the example of gauge theory above,
there is a natural geometrical way to include this collection of
fields in an action principle. Let us introduce a Euclidean simplex
$S$ of large dimension. The dimension of $S$ should be at least $N$
where $N$ is the number of smooth regions. Let $\{t^i\}$ be the
co-ordinates on the simplex: $\sum_i t^i = 1$, $t^i> 0$. We define
the linear combination of spin connections and also of vielbeins:
\begin{gather}
\bo := \sum_{i=1}^N t^i \omega_i\, ,\qquad \bE:= \sum_{i=1}^N t^i
E_i\, .
\end{gather}
Under local Lorentz transformations (which do not depend on $t$)
$\bo$ transforms as a connection and $\bE$ transforms as a vector.
i.e. $\bo \to \Lambda^{-1}\bo \Lambda + \Lambda^{-1}d\Lambda$, $E
\to  E \Lambda$ for $\Lambda = \Lambda(x)$. We define
\begin{gather}
\bO := d\bo + \bo \bo, \quad \bT := d\bE + \bo \bE, \quad \bD := d +
[\bo, \cdots],
\end{gather}
respectively the curvature, torsion and covariant derivative induced
on $M$ by $\bE$ and $\bo$. Using the exterior derivative on $S$,
denoted by $\delta$ we also define:
\begin{gather}
 {\cal F} : = \delta \bo +\bO ,\quad
 {\cal T} : = \delta \bE + \bT,\quad
 \nabla : = \delta + \bD,
\end{gather}
respectively the ``curvature", ``torsion" and ``covariant
derivative" induced onto the space $S \times M$. Let us introduce
the differential form, which we will call the \emph{secondary
Lagrangian}:
\begin{gather}\label{L_bar}
 {\cal K}:= \langle {\cal F}^p {\cal T}^q \bE^r \rangle.
\end{gather}
It can be expanded as a sum of differential forms of order $( p,
D-p)$ in $(dt, dx)$, each of which is invariant under local Lorentz
transformations of the form $\Lambda(x)$. Finally, we note that the
curvature and torsion on $S\times M$ obey the usual Bianchi
Identities $\nabla {\cal F}^{ab}= 0$ and $\nabla {\cal T}^a = {\cal
F}^a_{\ b}\bE^b$.
Also, by the invariance of the secondary Lagrangian, we have $
(\delta +d) {\cal K} = \nabla {\cal K}$.

We can then define the following action, which we call the
\emph{secondary action}:
\begin{gather}
 I[\{ E_i, \omega_i\}] := \sum_A \int_{\Sigma_A} \int_{c^A} {\cal K}
\end{gather}
where $\Sigma_A$ are sumbanifolds of $M$ of dimension $D-p$ and
$c^A$ are corresponding submanifolds of $S$ of dimension $p$. The
dimension $p$ is summed over. The domain of integation is a kind of
composite space involving submanifolds of $M$ and cells in $S$ which
we will call the \emph{secondary manifold}, $W:= \sum_A c^A
\times\Sigma_A$. The secondary action contains a sum of integrals
over the bulk regions $\sum_i \int_{\Sigma_i} \langle (\Omega_i)^p
(T_i)^q (E_i)^r\rangle = \sum_i \int_{\Sigma_i} \cL (E_i, \omega_i)$
plus surface terms at the hypersurfaces and their intersections. The
surface terms are functionals of the collection $\{E_i, \omega_i\}$
associated with all the bulk regions which meet at the intersection.
As such, these surface terms are a good candidate to be the
generalisation of the term (\ref{JumpK}) describing the
distributional geometry. Such an action was first suggested for
Lovelock gravity in Ref. \cite{Gravanis-04} and was shown to
correctly describe the distributional geometry. Here we wish to take
the treatment further, applying it to the case of more general
theories of gravity with torsion. Also we shall study the field
equations as well as the action.

In the next section, we discuss more explicitly the construction of
the secondary manifold and the secondary action, as well as
introducing a smoothing process showing how the secondary action may
arise as the limit of an action for a single smooth geometry.

\subsection{Cell complexes}

Let's say we have a geometry given by vielbein $E$ and spin
connection $\omega$ on $M$, which are piecewise smooth. The cell
structure is chosen so that, in the interior of each bulk cell
$\Sigma_i$, the fields are smooth: e.g. $E = E_i$, a smooth function
and $\omega = \omega_i$. On the lower dimensional cells the fields
may be discontinuous.

By smooth, we mean that the field is sufficiently differentiable.
For example, in Einstein-Cartan gravity, the vielbein and spin
connection should be $C^1$ in the interior of the bulk cells. Since
the curvature and torsion tensors contain first derivatives of these
quantities, they are continuous tensor fields in the interior of
each bulk cell, but may be a distribution valued tensor on some
submanifolds of lowed dimension. This would describe a collection of
thin domain walls, cosmic strings etc.

\begin{definition}[Cell complex on $M$]
Let $M$ be a manifold, of dimension $D$, composed of closed cells
$\Sigma$. The cells are of various co-dimension $p$ i.e. dimension
$D-p$. We label the cells of co-dimension zero (which we call bulk
cells) $ \Sigma_i$, $i= 1,\dots N $. A general cell is labelled
$\Sigma_A$ where $A$ is an abstract index uniquely identifying each
cell. The collection of cells $\{\Sigma_A\}$ cover the whole
manifold so that $M = \bigcup_A \Sigma_A$. For the moment, we shall
assume that $M$ is without boundary.

Each cell has an intrinsic orientation. The boundary of a cell of
codimension $p$ will be a linear combination of cells of
co-dimension $p+1$:
\begin{equation}\label{oriented}
\partial \Sigma_B = \sum_A\epsilon(A,B) \Sigma_A,
\end{equation}
where the numbers $\epsilon(A,B)$ are depending on the orientation
induced on $\Sigma_A$ by $\Sigma_B$. [The boundary of $\Sigma_B$ as
a set contains all the cells $\Sigma_C$ in $\partial \Sigma_A$ and
all the cells in $\partial \Sigma_C$ and so on.] The numbers
$\epsilon(A,B)$, which we shall call the incidence numbers, satisfy
\begin{equation}\label{indidence numbers constraint}
\sum_B \epsilon(A,B) \epsilon(B,C)=0\ .
\end{equation}
\end{definition}

\begin{definition}[Abstract dual cells]\label{dual cells}
The dual cell complex is a space
which is built
according to the following rules:
\\
 i) Corresponding to each bulk cell $\Sigma_i$,
we assign a dual cell $c^i$ which is an abstract point;
\\
ii) For each cell $\Sigma_A$ of co-dimension $p$, there is a
corresponding dual cell $c^A$ which is a smooth manifold of
dimension $p$, such that:
\begin{equation}\label{dual_definition}
 c^B \subset \partial c^A \quad \Leftrightarrow\quad \Sigma_A
\subset \partial \Sigma_B.
\end{equation}
\\
iii) The dual cell complex is homeomorphic to a Euclidean or
geometric simplicial complex~\footnote{This requirement restricts
appropriately the too great generality in the type of complexes
implied otherwise by our definitions. The problem has to do with the
way cells are attached. Our cell complexes are CW
complexes~\cite{Massey} but we do not allow every kind of CW
complex. For a general CW complex the fields across cell boundaries
would then be related by the non-trivial attaching maps. It is not
clear how to formulate the problem in such a case.}. We may imagine
the simplicial complex `living' in large enough simplex $S$, which
we have already mentioned above.

The relative orientations of the $c^A$ are specified by the boundary
map of the cell complex. Cells $c^A$ are chosen to be co-oriented
with respect to the $\Sigma_A$ in the following sense:
\begin{equation}\label{co_oriented}
\partial c^A = \sum_B
\epsilon(A,B)\, c^B (-1)^{\text{dim} (c^B) + 1}\, ,
\end{equation}
where the numbers $\epsilon(A,B)$ are defined by equation
(\ref{oriented}).
\end{definition}
The dual cell complex is a cell complex with incidence numbers
$\epsilon(A,B)\, (-1)^{\text{dim} (c^B) + 1}$.

We denote the two boundary maps (operators) acting on the two cell
complexes by the same symbol as no confusion arises. It can be
thought of as the formal sum of these maps which acts on the product
cell complex which we define below. It will mainly be used that way.
$\partial$ takes into account orientations and is required to
satisfy $\partial\partial=0$. Relation (\ref{indidence numbers
constraint}) was introduced to guaranty that.

For the time being we will think in terms of the large simplex $S$
describing our cell complex in there. Later on we will use the
complexes in a more direct way.

\begin{figure}[h]
  \input{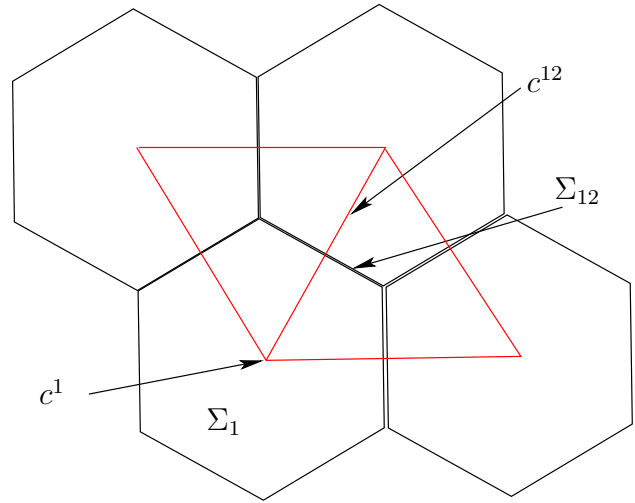}\\
  \caption{{\small A part of a two dimensional complex is shown.
  The dual complex is drawn over the top. Selected cells are labelled.
$c^1$
  is a vertex of the dual complex, and is dual to
  the bulk cell $\Sigma_1$.}}\label{dual_complex}
\end{figure}

The dual lattice $\{ c^A\}$ in $S$ defined above is homeomorphic to
the standard dual lattice in $M$ which one obtains by placing
vertices in the center of each $\Sigma_i$ and then joining together
with dimension 1 cells to form the skeleton etc. \cite{Book}. This
is illustrated in fig. \ref{dual_complex}.

\begin{definition}[Other cells in $S$]
A general vertex in $S$ is labelled $\langle \alpha \rangle$. We
shall include vertices, $\alpha \notin 1,\dots, N$, which are not
dual to any of the bulk cells on $M$. We include these because they
are useful for keeping track of other fields defined over the cells
or the whole manifold.
\end{definition}

\begin{figure}[h]
  \includegraphics[width=.49\textwidth]{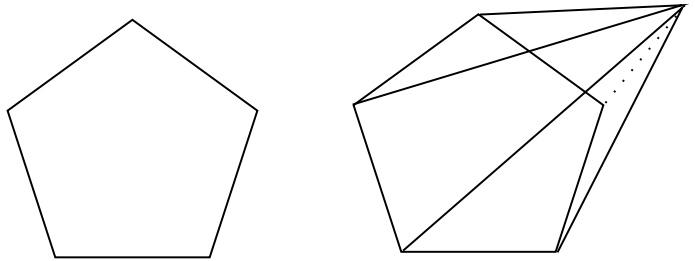}\\
  \caption{\small Left: A two-dimensional cell, $c$. Right: The cone
product of $c$ with another vertex.} \label{homotopy_product_fig}
\end{figure}

\begin{definition}[Cone product]
The cone product of a vertex with a dual cell is defined as follows:
Let $c$ be a dual cell of dimension $p$. Then $C(\alpha,c)$ is a
cell of dimension $p+1$ obtained by joining all vertices of $c$ to
the vertex $\alpha$, with orientation such that:
\begin{gather}\label{partial_homotopy_product}
\partial C(\alpha, c) = c - C(\alpha, \partial c) - \langle\alpha\rangle
\delta_{\!p}.
\end{gather}
The Kronecker delta $\delta_p  =1$ if dim $c$ $=0$ and is zero
otherwise. We also defined $C(\alpha,\varnothing)
\equiv\varnothing$.
\end{definition}
The cone product is illustrated by fig. \ref{homotopy_product_fig}.


\subsection{Secondary action and smoothing theorem}\label{Secondary action and smoothing theorem}

\begin{definition}[Product of cell complexes]\label{product of
cells} Let $\{\Sigma\}$ and $\{c\}$ be two cell complexes. Their
product $\{c \times \Sigma\}$, is the set of all the Cartesian
products $c^A\times \Sigma_B$ with the boundary rule
\begin{equation}\label{boundary of product}
\partial (c^A\times
\Sigma_B)=\partial c^A\times \Sigma_B+(-1)^{\text{dim}(c^A)}
c^A\times
\partial\Sigma_B\,.
\end{equation}
\end{definition}
Clearly the product cell complexes is a cell complex as the boundary
rule can be put in the form (\ref{oriented}) for appropriate
incidence numbers.

In a product of cells it is not always necessary both cells belong
in cell complexes. The following more general structure is also
useful.
\begin{definition}[Cell complex with cell-valued coefficients]
\label{Cell complex with cell-valued coefficients} Let a cell
complex $\{\Sigma_A\}$ and a set of cells $\{d^A\}$ which is, or is
a subset of, a cell complex $c$. The rule (\ref{boundary of
product}) defines a cell complex of $\Sigma$ cells with $c$-valued
coefficients.
\end{definition}
Of course a product $\{c \times \Sigma\}$ can always be thought of
as $\Sigma$ cell complex with $c$-valued coefficients, or a $c$ cell
complex with $\Sigma$-valued coefficients.

\begin{convention}
Sums of cells of a cell complex will be called \textit{chains}.
Therefore any cell may also be thought of as a chain. Also, as the
cell complex is homeomorphic to a geometric complex, any cell is
homoemorphic to a chain of simplices in the geometric complex. We
will treat cells $c^A$ as chains also that way.

We will alternatively use the equivalent notations: $|c| \equiv
\text{dim}(c)$ and $|A| \equiv \text{dim}(c^A) $.
\end{convention}

We now formally define the secondary manifold mentioned above.
\begin{definition}[Secondary manifold
$W$]\label{Secondary_Manifold_def} Let $\Sigma_A$ be a cell complex
and its dual complex $c^A$. Let a chain in their product complex
defined by:
\begin{gather}
    W := \sum_A c^A \times \Sigma_A.
\end{gather}
If $M$ is the manifold which is the union of the cells $\Sigma_A$
then $W$ is the called the \textit{secondary manifold}. Clearly $M$
and $W$ have the same dimension.
\end{definition}

The boundary of the secondary manifold is
\begin{equation}\label{partial_W}
\partial W = \varnothing.
\end{equation}
It is straightforward to derive this, making use of (\ref{oriented})
and (\ref{co_oriented}) to cancel $c\times \partial \Sigma$ terms
with $\partial c \times \Sigma$ terms.

\begin{definition}[Smoothing manifold $W^\alpha$]
Let $\langle \alpha\rangle$ be a vertex in $S$ which is \emph{not}
dual to any of the cells $\Sigma$. The smoothing manifold with
respect to $\alpha$ is:
\begin{gather}
    W^\alpha :=
    \sum_A C(\alpha,c^A) \times \Sigma_A.
\end{gather}
\end{definition}

We have defined $W$ as a chain, a formal sum over cells. However the
name secondary manifold is reasonable since $W$ can be given the
structure of a manifold under mild topological assumptions. For the
examples we will consider in this paper $W$ is a manifold, as is
$W^\alpha$.

The reason why $W^\alpha$ is called the smoothing manifold will
become apparent below. First we note that, using
(\ref{partial_homotopy_product})
and $\partial W = \varnothing$ we get
\begin{gather}\label{smoothing}
\partial W^\alpha = W - \langle\alpha\rangle \times M.
\end{gather}
The vertex $\langle \alpha\rangle$ is simply a point so
$\langle\alpha\rangle \times M$ is isomorphic to the spacetime
manifold. So the smoothing manifold is a co-bordism between $M$ and
the secondary manifold. It is natural that $\langle\alpha\rangle
\times M$ will arise in describing a single field over the whole of
$M$ whereas $W$ is suitable for describing a collection of fields
living in the bulk cells (and possibly also fields living only on
the cells of lower dimension). The function of the smoothing
manifold is to provide a link between the two situations.

As mentioned above, we shall mainly be interested in considering a
Lagrangian of the form ${\cal L}(E,\omega) = \langle \Omega ^p T^q
E^r\rangle$. More generally, we may consider a Lagrangian ${\cal
L}(\Psi)$ composed of a field or collection of fields, $\Psi$, and
its derivatives. Suppose then that the action which defines our
theory is:
\begin{gather}\label{Psi_action}
 I[\Psi] := \int_M {\cal
 L}(\Psi).
\end{gather}

\begin{definition}[The limiting process, the bulk fields $\{\Psi_i\}$]
Let $\Psi_\alpha$ be a sequence of smooth fields (or collections of
fields), parametrised by integer $\alpha$, which tend towards a
distributional field as $\alpha \to \infty$. We shall assume that
the distributional field defined by the limit is smooth in the
interior of each bulk cell and that it has at most a bounded
discontinuity across any cell of codimension $p$ for $p >0$.

On each bulk cell, $\Sigma_i$, we define the smooth field $\Psi_i$
as follows: $\lim_{\alpha \to \infty} (\Psi_\alpha - \Psi_i) = 0$
everywhere in the interior of $\Sigma_i$. We shall call $\Psi_i$
bulk fields.
\end{definition}

Note that the value of $\Psi_i$ is well defined at $\partial
\Sigma_i$ by continuity but the value of $\lim_{\alpha \to
\infty}\Psi_\alpha$ on $\partial \Sigma_i$ is in general undefined.

\begin{definition}[Secondary Lagrangian ${\cal
K}$]\label{Secondary_Lagrangian_def} Assume that the theory is
defined by the action (\ref{Psi_action}) for field(s) $\Psi$. Then
the secondary Lagrangian ${\cal K}$ is a $D$-form constructed from
the bulk fields $\Psi_i$ such that:
\\
 i) ${\cal K}$ is invariant, up to an exact part, with respect to diffeomorphisms of $M$ and gauge transformations
 which depend on the co-ordinates of $M$;
\\
 ii) The pullback of ${\cal K}$ onto $\langle \alpha \rangle \times M$ coincides with
  the original Lagrangian evaluated for the field
$\Psi_\alpha$. This implies:
\begin{gather}
 \int_{\langle \alpha \rangle \times M}{\cal K} = \int_M {\cal
 L}(\Psi_\alpha) = I[\Psi_\alpha]\, .
\end{gather}
\end{definition}
\begin{definition}[Secondary action]
 The secondary action is:
 \begin{equation}
  I[\{ \Psi_i\}] := \int_{W} {\cal K}\, .
 \end{equation}
 It is a functional of the bulk fields.
\end{definition}

We first use (\ref{smoothing}) to get:
\begin{gather*}
\int_W {\cal K} - \int_{\langle \alpha \times M\rangle } {\cal
  K}
= \int_{\partial W^{\alpha}} {\cal K}.
\end{gather*}
 Then using the above definitions we see that the left hand side is
 the difference between the secondary action and $I[\Psi_\alpha]$.
 Using Stokes' Theorem on the right hand side we get:
\begin{gather}\label{Nice_formula}
 I[\{\Psi_i\}] - I[\Psi_\alpha] = \int_{W^{\alpha}}(\delta + d) {\cal K}.
\end{gather}
We thus obtain the following result:
\begin{theorem}[Smoothing Theorem - manifold without boundary]
\label{smoothing theorem}
In the limit $\alpha \to \infty$ the action $I[\Psi_\alpha]$
converges unambiguously to the secondary action $I[\{\Psi_i\}]$ if
and only if
\begin{equation}\lim_{\alpha \to
\infty}\int_{W^\alpha}(\delta + d) {\cal K} = 0\, .
\end{equation}
Conversely, if the above condition is satisfied, the secondary
action can be approximated arbitrarily closely by the action
$I[\Psi_\alpha]$ for a smooth field $\Psi_\alpha$ defined over $M$.
\end{theorem}
If the above condition is satisfied for a certain class of fields,
the action is said to be \emph{smoothable}. Smoothability for a
manifold with boundary will be treated in section \ref{Manifolds
with boundary}.

\subsection{Finding the secondary Lagrangian}\label{Finding the secondary Lagrangian}

Now, it is not obvious that, for any given Lagrangian, a secondary
Lagrangian obeying properties i) and ii) of definition
\ref{Secondary_Lagrangian_def} exists. Essentially, the secondary
Lagrangian is a co-chain i.e. a linear map from the product complex
$\{c \times \Sigma\}$ to the real numbers, whose value on $W$ is the
secondary action. It is not obvious that the secondary action, that
is that such a co-chain, exists. Also if it does exist it is not
automatic to write it down in some general form. In this work we
give some examples where these are possible.

We have seen in section \ref{Idea_section} that, for the geometrical
actions constructed from vielbein $E^a$ and the curvature
$\Omega^{ab}$, one can surely construct functionals out of the
`Lorentz' tensors $\delta \bar E, \delta \bar\omega, \bar E,
\bar\Omega, \bar T$. All these are differential forms over $S \times
M$. One can even group them up to write down a secondary Lagrangian,
for example
\begin{gather*}
 \left\langle (\alpha_1 \delta \bo + \bO) \cdots (\alpha_p \delta
 \bo + \bO) (\beta_1 \delta \bar E + \bT) \cdots (\beta_q \delta \bar E + \bT)
 \bE^r \right\rangle\,,
\end{gather*}
for some numbers $\alpha_1,\dots,\beta_q$. The problem with this
general choice is diffeomorphism invariance.

Let $\xi$ be an arbitrary vector field in the tangent space of $M$.
The secondary actions are required to be invariant under the
(diffeomorphism) transformations it generates, essentially
infinitesimal change of coordinates. This invariance is elegantly
imposed using Cartan's identity for the Lie derivative:
$\pounds_\xi=i(\xi)d+d i(\xi)$, where $i(\cdot)$ the contraction
operator. As $\xi$ does not depend on the coordinates of the dual
space we can also write $\pounds_\xi=i(\xi)(\delta+d)+(\delta+d)
i(\xi)$. We used that $\delta i(\cdot)=-i(\cdot)\delta$. Then by
$\delta^{DIFF}_\xi \int_W {\cal K}=\int_W \pounds_\xi {\cal K}$ we
have:
\begin{equation}\label{}
\int_W i(\xi) (\delta +d) {\cal K}+\int_{\partial W} i(\xi) {\cal
K}=0\,.
\end{equation}
For $M$ without boundary we have $\partial W=\varnothing$, so the
second term vanishes. Therefore by the last formula we know how to
impose diffeomorphism invariance on the secondary action. We will
discuss it further later on but for the moment let's apply it to a
specific case.

Consider the simple example of a 2-dimensional manifold and the
secondary action
\begin{equation}\label{}
\int_W \langle \alpha_1 \delta \bar \omega+\bar \Omega \rangle =
\int_W \epsilon_{ab} \left(\alpha_1\delta \bar\omega+\bar
\Omega\right)^{ab}\,.
\end{equation}
If proven smoothable this should be proportional to the Euler number
of the 2-manifold, calculated by a discontinuous connection i.e.
non-smooth metric in its tangent bundle. But before everything we
should better check the invariance under those transformations
induced by a vector field. Note that $(\delta+\bar D)\delta \bar
\omega=\bar D \delta \bar\omega=-\delta \bar \Omega$ and
$(\delta+\bar D)\bar \Omega=\delta \bar\Omega$. So
\begin{equation}\label{}
\delta_\xi\int_W \langle \alpha_1 \delta \bar \omega+\bar \Omega
\rangle = (1-\alpha_1)\int_W i(\xi)\epsilon_{ab} \delta \bar
\Omega^{ab} \,.
\end{equation}
The r.h.s. vanishes only if the curvature is continuous or
$\alpha_1=1$. This generalizes to higher dimensional cases. It is
not possible to have the invariance as a general property of the
theory unless all those numbers are equal to 1.

The general linear combination $a_1 \delta \bar \omega+\bar \Omega$
transforms under diffeomorphisms in the same way as the curvature
$\mathcal{F}=\delta \bar \omega+\bar \Omega$ i.e. $\pounds_\xi
\mathcal{F}=i_\xi(\nabla \mathcal{F})+\nabla (i_\xi
\mathcal{F})-[i_\xi \bar \omega,\mathcal{F}]$, the difference being
that the curvature satisfies also the Bianchi identity $\nabla
\mathcal{F}=0$, as discussed in section \ref{Idea_section}, while
$\nabla(\alpha_1 \delta \bar \omega+\bar \Omega)=(1-\alpha_1)\delta
\bar \Omega$.

It therefore makes sense, as much as it is natural, to build
Lagrangians using the curvature $\mathcal{F}$ and torsion
$\mathcal{T}$ and in general geometric objects over $S\times M$, as
is for example the case of (\ref{L_bar}). Then the smoothing
manifold is a useful and elegant way to treat the limiting processes
for smoothing out a piecewise smooth geometry.

\section{Some applications of the smoothing theorem}

We have a single condition, that the exterior derivative of the
secondary action vanish on the smoothing manifold. It was shown in
ref. \cite{Gravanis-04} that this condition is satisfied for
Lovelock gravity in the second order formalism (connection is
determined from vielbein by the zero torsion constraint), provided
that the vielbein is continuous. In ref. \cite{Willison-04}, it was
argued that the condition is satisfied for a more general constraint
on the torsion. In the following we will see how these requirements
on torsion and continuity can be relaxed further.

\subsection{Einstein-Cartan gravity}\label{Einstein-Cartan gravity}

Perhaps the most relevant application of this geometrical
construction is when the Lagrangian in question is that of general
relativity in four dimensions. Thus we choose ${\cal L} = R
\sqrt{-g}\, d^4 x$. Even in this familiar case, there are some
surprises to be found. Writing the Lagrangian in terms of
differential forms we have:
\begin{gather}
 {\cal L}(E, \omega) = \frac{1}{2}\,
 \epsilon_{\hat{a}\hat{b}\hat{c}\hat{d}}\,  \Omega^{\hat{a}\hat{b}}
 \wedge E^{\hat{c}} \wedge E^{\hat{d}}\, .
\end{gather}
We shall assume that the vielbein $E^{\hat{a}}$ and spin connection
$\omega^{\hat{a}\hat{b}}$ are in principal independent. The field
equations in empty space would give vanishing torsion, fixing
$\omega^{\hat{a}\hat{b}}$ in terms of $E^{\hat{a}}$, but if there is
some matter which couples to the spin connection the torsion could
be non-vanishing. In particular, we will allow for the possibility
of some matter with spin located on a cell in $M$. This
generalisation of GR with torsion is normally known as
Einstein-Cartan theory.

We will consider a simple scenario: there is one hypersurface
dividing the space-time into two bulk regions. So let $M$ be a
4-dimensional space-time manifold. We assume that $\partial M$ is
far away and that the asymptotic conditions on the fields are such
that it can be ignored. $M$ has the following cell structure:
$\Sigma_1$ and $\Sigma_2$ are cells of co-dimension 0 (the bulk
regions) and $\Sigma_{12}$ is a cell of co-dimension 1, oriented
such that $\partial \Sigma_1 = -\partial \Sigma_2 = \Sigma_{12}$.
The corresponding dual cells are $c^1$, $c^2$ and $c^{12}$.
According to equation (\ref{co_oriented}), these have orientation
defined by $\partial c^{12} = c^2 -c^1$.

It is convenient to drop the indices and write e.g. $\epsilon(\Omega
E E) := \epsilon_{\hat{a}\hat{b}\hat{c}\hat{d}}\,
\Omega^{\hat{a}\hat{b}}\wedge E^{\hat{c}} \wedge E^{\hat{d}}$. When
all of the indices are contracted with the epsilon tensor there is
no ambiguity in this notation. The secondary Lagrangian is obtained
by replacing $\Omega$ with ${\cal F} = \bO + \delta \bo$ and $E$
with $\bE$ as defined in section \ref{Idea_section}.
\begin{gather}
{\cal K} = \frac{1}{2}\,\epsilon (\bO \bE \bE) +
\frac{1}{2}\,\epsilon(\delta \bo \bE\bE)\, .
\end{gather}
The secondary action is $I[E_1, \omega_1, E_2,  \omega_2] : = \int_W
{\cal K}$ where $W$ is the secondary manifold introduced in
definition \ref{Secondary_Manifold_def}. The secondary manifold and
the smoothing manifold $W^\alpha$ are illustrated in Fig.
\ref{S12Wfig}. After performing explicitly the integral over
$c^{12}$ we get:
\begin{gather}
I[E_1,\!\omega_1,\! E_2,\!  \omega_2] = \frac{1}{2}\int_{\Sigma_1}\!
\epsilon(\Omega_1 E_1E_1) + \frac{1}{2}\int_{\Sigma_2}\!
\epsilon(\Omega_2 E_2E_2) \nonumber
 \\  + \frac{1}{24} \int_{\Sigma_{12}} \epsilon\big(
(\omega_2-\omega_1)\{(E_2-E_1)^2 + 3(E_2+E_1)^2\}\big).
\end{gather}
It is clear that the secondary action describes some discontinuous
geometry where the vielbein and spin connection are double-valued at
$\Sigma_{12}$. Suppose that we have some sequence of smooth
geometries given by $E_\alpha$, $\omega_\alpha$ such that in the
limit $\alpha \to \infty$ they become discontinuous. The question
is, under what conditions does the action $I[E_\alpha,
\omega_\alpha] = \int_M {\cal L}(E_\alpha, \omega_\alpha)$ converge
to the secondary action? The answer, according to theorem
\ref{smoothing theorem}, is when the integral over $W^\alpha$ of
$(d+\delta)\cK$ vanishes. Expanding this in powers of $(dx,dt)$ we
get:
\begin{align*}
&\quad \text{terms}   & \text{order in} (dx, dt)   \\
  (d+\delta)\cK &= \epsilon(\bO \bT \bE)  & (5,0)\
  \\&\quad - \epsilon( \delta \bo \bT\bE) + \epsilon(\bO \delta \bE \bE)
&(4,1)\
  \\&\quad  - \epsilon( \delta \bo \delta \bE \bE) &(3,2).
\end{align*}
When restricted to $W^\alpha$, the first term contributes nothing.

The next pair of terms are evaluated on $C(\alpha, c^i) \times
\Sigma_i$. In the limit $\alpha \to \infty$ the test fields
$(E_\alpha, \omega_\alpha)$ must coincide with the fields
$(E_i,\omega_i)$ in the interior of region $\Sigma_i$. Hence $\delta
\bE$ and $\delta \bo$ $\to 0$ on $C(\alpha,c^i)\times \Sigma_i$ and
these two terms do not contribute.

The final term is more problematic. It is to be integrated over
$C(\alpha, c^{12})\times \Sigma_{12}$. Performing explicitly the
integral over the triangle $C(\alpha, c^{12})$ gives:
\begin{widetext}
\begin{equation*}
-\int_{\Sigma_{12}}\int_0^1 dt^1\int_0^{1-t^1} dt^2 \epsilon\big(
\{(\omega_1 - \omega_\alpha)(E_2 - E_\alpha) - \langle 1
\leftrightarrow 2\rangle \} \{t^1(E_1 - E_\alpha) +t^2(E_2 -
E_\alpha)+E_\alpha\}\big)
\end{equation*}
\begin{align}
 & = -\frac{1}{6}\int_{\Sigma_{12}}\epsilon\Big( \big\{
 \omega_\alpha (E_1- E_2) -(\omega_1 -\omega_2) E_\alpha
 + \omega_1 E_2 -\omega_2 E_1\big\} \big\{E_1+E_2 + E_\alpha\big\}
 \Big)\label{Bad_term}
\end{align}

\begin{figure}[h]
  \includegraphics[height=.99\textwidth, angle=90]{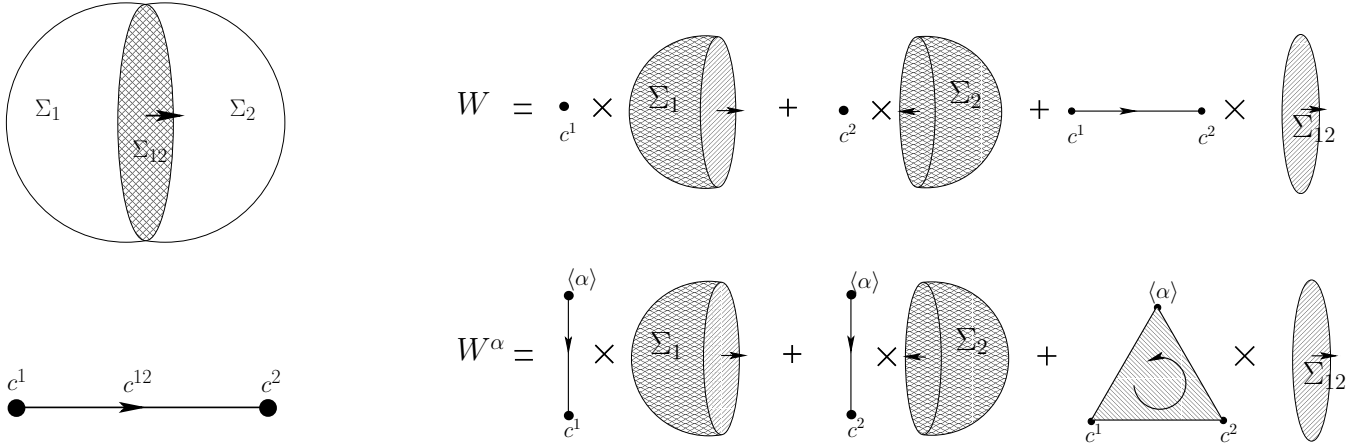}\\
  \caption{{\small Top left: The manifold $M$ is divided into bulk regions $\Sigma_1$ and $\Sigma_2$ by
  the hypersurface $\Sigma_{12}$. For convenience $M$ is shown as a compact ball
  but we should emphasize that we are ignoring the boundary $\partial M$ at the moment.
  Bottom left: the corresponding dual cells $c^1$, $c^2$ and $c^{12}$ with
  orientation induced by  the $\Sigma$'s.
  Top right: The secondary manifold of $M$ is $W =c^1\times \Sigma_1 +
   c^2 \times \Sigma_2 + c^{12} \times
   \Sigma_{12}$.
  Bottom right: The smoothing manifold $W^\alpha  = C(\alpha, c^1)\times \Sigma_1 +
   C(\alpha, c^2) \times \Sigma_2 + C(\alpha, c^{12}) \times
   \Sigma_{12}$.  }
  }\label{S12Wfig}
\end{figure}
\end{widetext} Requiring that this term vanish, there are two
possibilities:
\\
\textbf{1) Continuous vielbein, discontinuous spin connection:} If
$E_1|_{\Sigma_{12}} = E_2|_{\Sigma_{12}}$ then $\omega_\alpha$ drops
out of (\ref{Bad_term}). Furthermore, since the vielbein is
continous, the limiting value of the field $E_{\alpha}$ is well
defined on $\Sigma_{12}$ and is given by:
\begin{equation*}
\lim_{\alpha \to \infty} E_\alpha\big|_{\Sigma_{12}}=
E_1\big|_{\Sigma_{12}} = E_2\big|_{\Sigma_{12}}\, .
\end{equation*}
The problematic term (\ref{Bad_term}) vanishes in this case. So, by
the smoothing theorem, in the limit that $\omega_\alpha$ becomes
discontinuous, the action converges unambiguously to
$I[E_1,\!\omega_1,\! E_2,\! \omega_2]$, this being the sum of
Einstein-Hilbert terms in the bulk plus a surface term:
\begin{align}
I[E_1,\!\omega_1,\! E_2,\!  \omega_2] &=
\frac{1}{2}\int_{\Sigma_1}\! \epsilon(\Omega_1 E_1E_1) +
\frac{1}{2}\int_{\Sigma_2}\! \epsilon(\Omega_2 E_2E_2)\nonumber \\
 & \qquad +\frac{1}{2} \int_{\Sigma_{12}} \epsilon\big(
(\omega_2-\omega_1) E^2\big)\, .
\end{align}
In the absence of torsion, the continuity of the metric implies the
continuity of the tangential components of the spin connection. In
that case the surface term is the difference of the Gibbons-Hawking
term on each side of $\Sigma_{12}$, as expected.
\begin{align}
 I[g_1,g_2] &= \int_{\Sigma_1}\!  R_1 \sqrt{-g_1}\, d^4 x + \int_{\Sigma_2}\!
 R_2\sqrt{-g_2}\, d^4 x  \nonumber
  \\ & \qquad\mp 2 \int_{\Sigma_{12}}  (K_2 -K_1) \sqrt{\mp h}\, d^3 x\, .
\end{align}
\\
\textbf{2) Discontinuous vielbein, continuous spin connection:} If
$\omega_1|_{\Sigma_{12}} = \omega_2|_{\Sigma_{12}}$ then $E_\alpha$
drops out of (\ref{Bad_term}). The field $\omega_{\alpha}$ converges
to a well defined value on $\Sigma_{12}$:
\begin{equation*}
\lim_{\alpha \to \infty} \omega_\alpha\big|_{\Sigma_{12}}=
\omega_1\big|_{\Sigma_{12}} = \omega_2\big|_{\Sigma_{12}}\, .
\end{equation*}
In this case also, the problematic term (\ref{Bad_term}) vanishes,
in spite of the fact that the limiting value of the vielbein
$E_\alpha$ is undefined on $\Sigma_{12}$.

The secondary action has vanishing surface term on $\Sigma_{12}$:
\begin{align}
I[E_1,\!\omega_1,\! E_2,\!  \omega_2] &=
\frac{1}{2}\int_{\Sigma_1}\! \epsilon(\Omega_1 E_1E_1) +
\frac{1}{2}\int_{\Sigma_2}\! \epsilon(\Omega_2 E_2E_2)
\end{align}
The possibility of a well defined action for a discontinuous metric
is quite remarkable. This can only happen if there is torsion. Let
us introduce Gaussian normal co-ordinates $( \xi^\mu, z)$ where and
$\xi^\mu$ are tangential and $z$ is the normal co-ordinate, chosen
so that the direction of increasing $z$ points from $\Sigma_1$ to
$\Sigma_2$, with $\Sigma_{12}$ located at $z=0$. Then the allowed
distributional parts of the torsion are:
\begin{widetext}
\begin{align}
 T^a_{z \mu} & = \left[(E_2)^a_\mu - (E_1)^a_\mu\right] \delta(z) +
 \left(\omega^a_{\ b \, z}[(E_2)^b_\mu - (E_1)^b_\mu]
 -  \omega^a_{\ b \, \mu}[(E_2)^b_z - (E_1)^b_z] \right) H(z)+\nonumber
 \cdots\, ,
\\\nonumber\\
 T^a_{\mu\nu} &= \left(\omega^a_{\ b \, \mu}[(E_2)^b_\nu - (E_1)^b_\nu]
 -  \omega^a_{\ b \, \nu}[(E_2)^b_\mu - (E_1)^b_\mu] \right) H(z)+\nonumber
 \cdots \,.
\end{align}
\end{widetext}
These field configurations are very exotic- the metric is undefined
at $\Sigma_{12}$. However, the metric is defined as the limit of
smooth metrics and we have proven that it gives a well defined
action. If one is performing a path integral for a theory with
torsion, it seems that these field configurations should not be
excluded.
\\

In the case where both spin connection and vielbein are
discontinuous, then the term (\ref{Bad_term}) generally does not
vanish. Indeed, making no assumptions\footnote{ In reference
\cite{Giacomini-06} it was assumed that $E_\alpha$ and
$\omega_\alpha$ are an interpolation with some arbitrary
\emph{scalar} function between $E_1$ and $E_2$ and $\omega_1$ and
$\omega_2$. Under that assumption the weaker condition $\left.
(\omega_2 - \omega_1)^{[ab}\wedge (E_2 - E_1)^{c]}
\right|_{\Sigma_{12}}=0$ for the vanishing of (\ref{Bad_term}) can
be obtained. This assumption is reasonable if there is some physical
constraint which ensures that the discontinuous parts of $\omega$
and $E$ are orthogonal to each other at every stage of the limiting
process. In the present work we shall make no assumptions about the
limiting process.} about the limit of $E_\alpha$ and
$\omega_\alpha$, it is undefined. Therefore situations 1) and 2)
above exhaust the possibilities.

The method of the smoothing manifold has been used to find some new
results for non-smooth geometry in Einstein-Cartan gravity. However,
these results could certainly have been found using a less
sophisticated formalism. Our approach really comes into its own when
the action is nonlinear in the curvature. A good example is Lovelock
gravity, to which we shall now turn our attention.

\subsection{Lovelock-Cartan Gravity}\label{Lovelock-Cartan Gravity}
Lovelock gravity in $D$ space-time dimensions is defined by a
Lagrangian which is a sum of terms $\epsilon ( \Omega \cdots \Omega
E \cdots E)$. Let us just focus on a single term, polynomial of
order $p$ in the curvature tensor.
\begin{gather}
{\cal L} = \epsilon(\Omega^p E^q)
\end{gather}
with $2p+q = D$.

The secondary action is:
\begin{gather}
{\cal K} = \epsilon([\delta \bo +\bO]^p \bE^{q}).
\end{gather}
In order to apply the smoothing theorem, we will need the following
result:
\begin{equation}
(d+\delta){\cal K} =  q \sum_{s=-1}^{p-1} {\cal N}_s+
q\sum_{s=0}^{p} {\cal M}_s,
\end{equation}
where ${\cal N}_s$ and ${\cal M}_s$ are differential forms of order
$(s+1, D-s)$ in $(dt,dx)$ given by
\begin{align}
 {\cal N}_s &:= \frac{p!}{(s+1)!(p-s-1)!}\, \epsilon\left( (\delta \bo)^{s+1}
\bO^{p-s-1} \bT \bE^{q-1} \right)\, ,\nonumber
\\
{\cal M}_s & := \frac{p!}{s!(p-s)!}\, \epsilon\left( (\delta
\bo)^{s} \bO^{p-s} \delta \bE \bE^{q-1} \right)\, .
\end{align}
Let us analyse the behavior of the integral over the smoothing
manifold. The integral is a sum of terms of the form:
\begin{gather}
 q \int_{C(\alpha, c)}  \int_{\Sigma} {\cal N}_s+{\cal M}_s\, ,
\end{gather}
where $\Sigma$ is a cell of codimension $s$. There are two generic
ways to make the integral vanish:
\\\\
\textbf{Case 1) Continuous vielbein, zero torsion, discontinuous
spin connection:} The cell $\Sigma$ is the intersection of a
collection of $n$ bulk cells $\{\Sigma_i| i \in (1, \dots , n)\}$.
Suppose that the intrinsic veilbein $E\big|_{\Sigma}$ on $\Sigma$ is
well defined i.e. $E_i \big|_{\Sigma} = E \big|_{\Sigma}\ \forall\
i\in (1, \dots n)$. The pullback of $E_\alpha$ therefore converges
to a well defined value $\lim_{\alpha \to \infty}
E_\alpha\big|_{\Sigma} =  E \big|_{\Sigma}$. This implies that the
contribution of ${\cal M}_s$, being proportional to $\delta \bar E$,
vanishes.

Now we observe that $\bar{T} = \sum_i t^i T_i + \sum_{i,j} t^i t^j
\omega_i (E_j-E_i)$. The first term vanishes by zero torsion and the
second by continuity of the vielbein. Therefore the contribution of
${\cal N}_s$, being always proportional to $\bar T$, vanishes.
\\\\
\textbf{Case 2) Continuous spin connection, discontinuous vielbein:}
In this case the pullback of $\delta \bar \omega$ vaishes. The only
terms which survive are ${\cal N}_{-1}$ and ${\cal M}_{1}$. ${\cal
N}_{-1}$ is a form of order $(0, D+1)$ so its pull-back identically
vanishes. ${\cal M}_{1}$ is of order $(1, D)$ so it is to be
integrated over $C(\alpha, c^i)\times \Sigma_i$ where $\Sigma_i$ is
a bulk cell. But, by construction $E_\alpha$ coincides with $E_i$ in
the bulk so the pull back of $\delta \bar E$ onto $C(\alpha, c^i)$,
and thus also ${\cal M}_{1}$, vanishes.
\\

The above two cases are the generic ways to satisfy the smoothing
theorem. There may also be some special cases depending on the
dimensionality of the cells at which the discontinuity occurs and on
which Lovelock terms are present in the action.

Note that our method applies to situations where the spin connection
is at worst $C^{1-}$ so solutions with solid angle defects of higher
codimension~\cite{Charmousis:2005ey}\cite{Zegers-08} fall outside of
our analysis.

\section{Homotopy and `renormalization'}

Up to this point we have presented some answers to questions posed
at the beginning of our work, but various issues remain still open.
Is the secondary manifold $W$ in any sense unique and what does it
mean if it is not? What happens if the manifold $M$ has a boundary?
What is the relation between the smoothability of the action and
that of the equations of motion? In this and the next section we
discuss these matters and their implications.

\subsection{The dual complex as a geometric complex}

Before going to analyze these issues we would like to discuss a bit
our requirement that the dual complex be homoemorphic to a geometric
simplicial complex. In particular to discuss certain reasonable
configurations which have to be excluded from a general formulation
because they bring bad company along with them. Being
half-justifiably sacrificed, they deserve now an honorary mention.

\begin{figure}[h]
  \includegraphics[height=0.30\textwidth, angle=0]{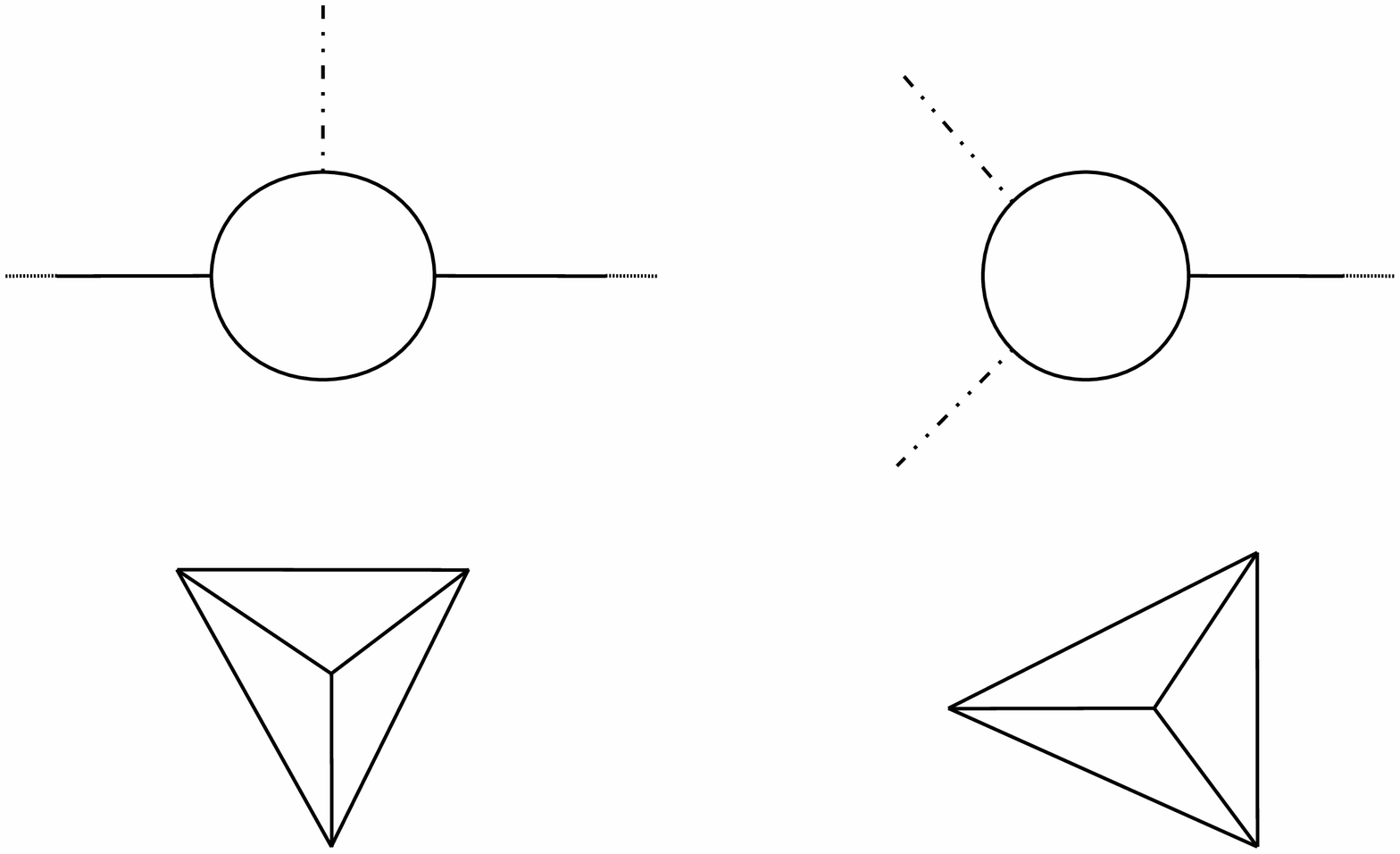}\\
  \caption{{\small Upper left: `Self-energy',
  Upper right: `Tadpole'. Bottom: The respective geometric dual complexes.}
  }\label{tadpole}
\end{figure}

Consider the configurations whose cross-sections are shown in the
figure~\ref{tadpole}: due to their similarity to Feynman graphs we
call them self-energy and tadpole graphs (they are given by the
continuous lines). These are reasonable intersections because, i)
locally, they look everywhere just fine and our considerations start
from local features, ii) there is a way to get them from
intersections with a geometric dual complex (as shown by the dashed
lines and we explain below). This latter fact is also a difficulty.

Self-energy is the better character of the two. It has this feature:
the boundary map $\partial$ acting on the $\Sigma$ cell complex
gives zero only acting at the highest co-dimension cells (one may
say its kernel is `trivial'). This is a good postulate for the cell
complex but excludes the tadpole, which as said already is fairly
reasonable. Put differently, an intersection like the tadpole
involves cells that have been identified along their boundary to
some extent.

Practically the problem is that if we apply the rules
(\ref{oriented}) and (\ref{co_oriented}) the thing works but we
obtain dual cell complexes that look strange. We want to express the
Lagrangian terms of each $\Sigma$ cell as integrals over plain
Euclidean simplices. Perhaps not surprisingly the problem can be
fixed: One way to look at it is that there are not enough bulk
regions in these configurations. So we may imagine, or start with
and then set to zero, new discontinuities (dashed lines) subdividing
the existing bulk regions and their boundaries. Then, as shown in
figure~\ref{tadpole}, the dual complex is a geometric complex.
Integrating the secondary Lagrangian over it, certain cells will
simply not contribute in the action as the fields are continuous
across the respective $\Sigma$ cells.

So every time we have mentioned `bulk cells' above, it could mean
bulk cells obtained by subdividing the initial bulk cells of a
configuration in a way to end up with geometric dual space. But if
we want to make all the distinction of what we can and cannot
describe at the level of the cell complexes, here lies the
difficulty. Imagine a single half-infinite hypersurface in spacetime
i.e. the connection is discontinuous only across. (Presumably this
is like a tadpole with a collapsed `circle'.) We can now imagine two
other (`dashed') half-infinite hypersurfaces forming a 3-way
intersection with it, a perfectly legitimate configuration. For many
Lagrangians (e.g. Lovelock gravities) we will find no term for the
co-dimension 2 cell this way. But we actually have a localized
holonomy here and there must certainly be such a term. So at this
point we think we have honored the victims enough. Empirically they
are of course acceptable. We will not try to refine further the cell
complex postulates in this work.

\subsection{The homology class of the secondary manifold}
Let us now turn to our questions at beginning of this section. First
of all, if for some configuration we can manufacture a manifold $W$
and prove smoothability, criterion \ref{smoothing theorem}, then we
definitely have a well-defined action for the distributional fields,
the secondary action $\int_W {\cal K}$. We may now explore the
following: The secondary manifold has been defined essentially
through the boundary relations (\ref{co_oriented}). They are chosen
so that if $M$ has no boundary then $\partial W=\varnothing$; that
is \emph{all} they contain. Such a condition does not define the
secondary manifold uniquely; the transformation $W \to W+\partial Y$
preserves the condition. Therefore $W+\partial Y$ can also be picked
as the secondary manifold, for any $Y$ such that $\partial Y$ and
$\partial \partial Y=\varnothing$ make sense. That is any given $W$
is a representative of a \emph{homology} class defined as the
equivalence class  $W \sim W+ \partial Y$. Fix a cell complex
$\{\Sigma_A\}$. Recalling the definition \ref{Cell complex with
cell-valued coefficients}, the general$D$-dimensional chain of such
a cell complex reads\begin{equation}Y=\sum_A d^A \times \Sigma_A\,
.\end{equation}Each $d^A$ is an arbitrary cell of dimension $|A|+1$
homeomorphic to chains of that dimension on a simplex, $S$, of very
large dimension. Recall that $|A|$ is the co-dimension of
$\Sigma_A$. [The set of chains of $S$ is the large cell complex
which `accommodates' all our cells and complexes.]

Then\begin{gather}\label{hatW}W+\partial Y= \sum_A \hat{c}^A \times
\Sigma_A\,,\end{gather}
where\begin{gather}\label{d-trans}\hat{c}^A=c^A +\partial d^A +
\sum_B (-1)^{|B|+1}\, \epsilon(A,B)\,d^B\,.\end{gather}Using $\sum_C
\epsilon(A,C)\epsilon(C,B)=0$ one finds that the cells $\hat{c}^A$
do indeed satisfy the same relations as the cells
$c^A$:\begin{equation}\label{co_oriented hat}\partial \hat{c}^A =
\sum_B(-1)^{|B| + 1}\,\epsilon(A,B)\, \hat{c}^B
\,,\end{equation}i.e. the incidence numbers $\epsilon(A,B)$
appearing here and in (\ref{co_oriented}) are the same. Thus we
have: the cells $c^A$ and $\hat{c}^A$ related by (\ref{d-trans}) are
\emph{equivalent choices} for dual cells. That is, if $\{\Sigma_A\}$
is a given cell complex, then $\{c^A\}$ and $\{\hat{c}^A\}$ are
equivalent choices for the cell complex dual to $\{\Sigma_A\}$. Thus
we arrive at the following\begin{definition}The set of all
homomorphic cell complexes, i.e. complexes with the same incidence
numbers, will be called a cell structure. A given cell complex is a
representative of the cell structure. [When no confusion arises we
denote the structure by one representative.]\end{definition}Using
this terminology we may say that a cell complex $\{\Sigma_A\}$
defines a dual cell \emph{structure}, not a single dual cell
complex. It turns out that that the cell structure can be viewed
also as a \emph{homotopy} class. From (\ref{co_oriented hat}) we
have that a linear combination of the cells $c^A$ becomes a linear
combination of $\hat c^A$ if induced by the same linear combination
of $d^A$. That is, an operator $\mathbf{P}$ defined by
$d^A=\mathbf{P}c^A$ is a linear operator acting on cells viewed as
chains. Then (\ref{d-trans}) tells us
that\begin{equation}\label{chain_homotopy}\hat{c}^A=c^A+\partial
\mathbf{P} c^A+ \mathbf{P} \partial c^A\,.\end{equation}In topology
chains $c^A$ and $\hat{c}^A$ related this way are said to be
chain-homotopic and the linear map $\mathbf{P}$ a
chain-homotopy~\cite{Hatcher}. Thus $\mathbf{P}$ moves the dual
cells around in the dual cell structure. Now any two elements in the
structure are not obviously homotopic. On the other hand their
defining relation (\ref{co_oriented}) is linear. That is the
structure is a linear space. There is always a linear map $\rho$
between any two cell complexes. Using linearity it is
straightforward to show via (\ref{co_oriented}) that
$\partial\,\rho=\rho\,\partial$. That is, $\rho$ is a chain-map.
Thus any two elements in the structure are related by a chain map.
Chain-maps fall into homotopy classes. Let $\mathbf{P}$ be a linear
map from cells of dimension $|A|$ to ones of dimension $|A|+1$. Then
$\partial\mathbf{P}+\mathbf{P}\partial$ is linear and commutes with
$\partial$ (by $\partial\partial=0$). Thus $\rho$ falls into some
equivalence class $\{\rho
\thicksim\rho+\partial\mathbf{P}+\mathbf{P}\partial\}$. That is any
chain-map belongs into an equivalence class of chain-homotopic maps.
In other words, the structure is in general a disjoint union of
chain-homotopic classes of cells.

There is a slightly different kind of homotopy operator which can be
naturally put on the same footing as $\mathbf{P}$. It is the cone
operator whose action is defined by
(\ref{partial_homotopy_product}). A cell complex which contains only
a vertex, $\{\langle *\rangle\}$, satisfies any given boundary rules
(\ref{co_oriented}), by $\partial \langle *\rangle=0$. That is
$\{\langle *\rangle\}$ belongs to any given dual cell structure. The
cone operator $C(*,\cdot)$ serves as a homotopy between $*$ and
other cell complexes. Given a cell $c^A$ and a vertex $*$, is it
always possible to form the cone $C(*,c^A)$? The answer is yes: all
our complexes essentially reside in the infinite dimensional
geometric simplex $S$. This is a contractible space therefore
homotopy, and homology as well, are trivial. This reflects the
properties we require at the level of fields: we want to approximate
in a continuous manner piece-wise smooth configurations of
$\{\Psi_i\}$ by single smooth fields $\Psi_\alpha$.
\begin{corollary} There is a single chain-homotopy class in the dual
cell structure. I.e. the structure is a homotopy class
itself.\end{corollary}We shall use this fact repeatedly in what
follows. Define now the action of $\mathbf{P}$ on chains of the
product complex $\{c \times \Sigma\}$, or more generally on a cell
complex of $\Sigma$ cells with dual-cell-valued coefficients, as a
linear operator acting by $\mathbf{P}(c^A \times
\Sigma_B):=(\mathbf{P}c^A)\times \Sigma_B$. By this definition one
finds that\begin{equation}\label{}(\partial
\mathbf{P}+\mathbf{P}\partial)(c^A \times\Sigma_B)=((\partial
\mathbf{P}+\mathbf{P}\partial)c^A)
\times\Sigma_B\,.\end{equation}That is $\mathbf{P}$ naturally
extends to a homotopy of chains on a product complex, thus we will
not use a different symbol for that extension. Then we have that the
chain $\hat{W} \equiv \sum_A \hat{c}^A \times\Sigma_A$ of
(\ref{hatW}) can be written
as\begin{equation}\label{}\hat{W}=W+\partial(\mathbf{P}W)+\mathbf{P}(\partial
W)\,.\end{equation} Recall that for $\partial M = \varnothing$ we
have $\partial W= \varnothing$ and therefore: $\mathbf{P}$ moves the
secondary manifold around in a \emph{homology
class}.\begin{definition}[Secondary manifold and its
class]\label{secondary manifold class} Let $\{\Sigma_A\}$ be a cell
complex whose union is a manifold $M$ without boundary. The
secondary manifold is a representative of a homology class (the
\textit{secondary manifold class} $\mathcal{W}_\Sigma$) of chains in
the $\Sigma$ cell complex with cell-valued coefficients in the dual
cell structure.\end{definition}That is, the secondary manifold is
one element of a homology class which is itself a homotopy invariant
(with respect to the dual space). By the previous corollary the dual
cell structure is unique. That is the class $\mathcal{W}_\Sigma$ is
\textit{unique}. This is our answer to the question of uniqueness of
the secondary manifold. Now this is a general mathematical result. A
question which arises
is:\begin{question}\label{secondary_class_question}What are all the
different representatives of the secondary manifold class
$\mathcal{W}_\Sigma$?\end{question}We take on answering that
question after we digress to deal with the case of $M$ with
boundary.

\subsection{Manifolds with boundary}
\label{Manifolds with boundary}

Let now $M=\bigcup_A \Sigma_A$ possess a boundary. Then some of the
cells will have a part of their boundary contained in $\partial M$.
Therefore we will generalize the boundary rule on cells as
\begin{equation}\label{}
\partial \Sigma_B=\sum_A \epsilon(A,B) \Sigma_A+ X_B\,,
\end{equation}
where $X_B=\partial \Sigma_B \cap \partial M$. Of course for some of
the values of the label these cells are empty.

$\partial \partial=0$ holds if and only if $\sum_A \epsilon(A,B)
X_A+\partial X_B=0$. This means that the boundary cells form a cell
complex with incidence numbers $-\epsilon(A,B)$. Thinking
intrinsically of the boundary $\partial M$ as a separate manifold,
the cells dual to $\{X_A\}$ obey
\begin{equation}\label{bdy co-oriented}
\partial c_{\|}^A=\sum_B (-1)^{|B|}\,\epsilon(A,B)\, c_{\|}^B\,.
\end{equation}
($|A|$ correctly gives the dimension of $c^A_{\|}$ and the
co-dimension of $\Sigma_A$.) Thus we deal with the boundary dual
cell structure.

The relation (\ref{bdy co-oriented}) is satisfied also by the cells
$(-1)^{|A|}\, c^A$. That is, these cells can be representatives of
the boundary dual cell structure, one may say `extrinsic' ones. If
it is unique, which we assume, then the structures $\{c^A_{\|}\}$
and $\{(-1)^{|A|}\, c^A\}$ coincide. Thus their cells are homotopic.
In other words there exists an operator $\mathbf{P}_{\|}$ such that
$\partial \mathbf{P}_{\|} c^A=(-1)^{|A|}\,
c_{\|}^A-c^A-\mathbf{P}_{\|}
\partial c^A$.

The cells $\mathbf{P}_{\|}((-1)^{|A|}c^A)$ are interpolations
between the `intrinsic' and `extrinsic' dual cells for $X_A$ as
cells in $\partial M$. They also have the right dimension and
boundary rule for dual cells of $X_A$ as cells in $M$.

On the direct sum of cell and boundary cell complexes (with
cell-valued coefficients) define the $D$-chain
\begin{equation}\label{full_W}
W:=\sum_A \left\{ c^A \times \Sigma_A+ (-1)^{|A|} \mathbf{P}_{\|}
c^A \times X_A \right\}\,.
\end{equation}

It will be useful to note the usual boundary rule
(\ref{co_oriented}) of the dual cells reads also $(-1)^{|A|}\partial
c^A=\sum_B \epsilon(A,B)\, c^B$. We may use this to show that
\begin{equation}\label{bdy of W}
\partial W=\sum_A c_{\|}^A \times X_A \,.
\end{equation}
As $\bigcup_A X_A=\partial M$ the manifold on the r.h.s. is nothing
but a secondary manifold associated with the cell structure in
$\partial M$.

The reason why the secondary manifold is defined by the chain $W$
above, is probably better understood by considering the action of
the cone-product which served in providing the right formulas to
state the `smoothing theorem' \ref{smoothing theorem} and also
discussed in the previous section. The cone-product is a
chain-homotopy $\mathbf{P}=C(*,\cdot)$ between any cell and a vertex
$\langle * \rangle$. By its very properties and the previous result
we find:
\begin{equation}\label{}
\partial C(*,W)=W-\left\{\sum_i \langle*\rangle \times
\Sigma_i+\sum_A C(*,c^A_{\|}) \times X_A \right\}\,.
\end{equation}
$\sum_i \Sigma_i=M$. The single vertex $\langle * \rangle$
represents smooth fields over $M$. In the presence of a boundary
$\partial M$ the action is defined by integrating the secondary
Lagrangian over the manifold in the brackets. That, for example,
reproduces the York (or Gibbons-Hawking) action of general
relativity. Thus $C(*,W)$ is again the smoothing manifold,
generalizing the discussion of section \ref{Secondary action and
smoothing theorem} in the presence of a boundary.

Consider now the action of an arbitrary homotopy on the chains
(\ref{full_W}). This involves a product of homotopy operators. It
will be convenient at this point to get more geometric. If a chain
$c$ is homotopic to a chain $\hat c$ via a homotopy $\mathbf{P}$,
relation (\ref{chain_homotopy}), we may define their \emph{prism
product} which is the cell $P(\hat c, c):=\mathbf{P}c$. From
(\ref{chain_homotopy}) we have that $\partial P(\hat c,c)=\hat c - c
-P(\partial \hat c, \partial c)$. Geometrically this makes sense as
being the boundary of the prism $P(\hat c, c)$ with $\hat c$ and $c$
as `top' and 'bottom' sides.

\begin{figure}[h]\vspace{-.0in}
  \begin{center}
  \includegraphics[width=0.45\textwidth,
  angle=0]{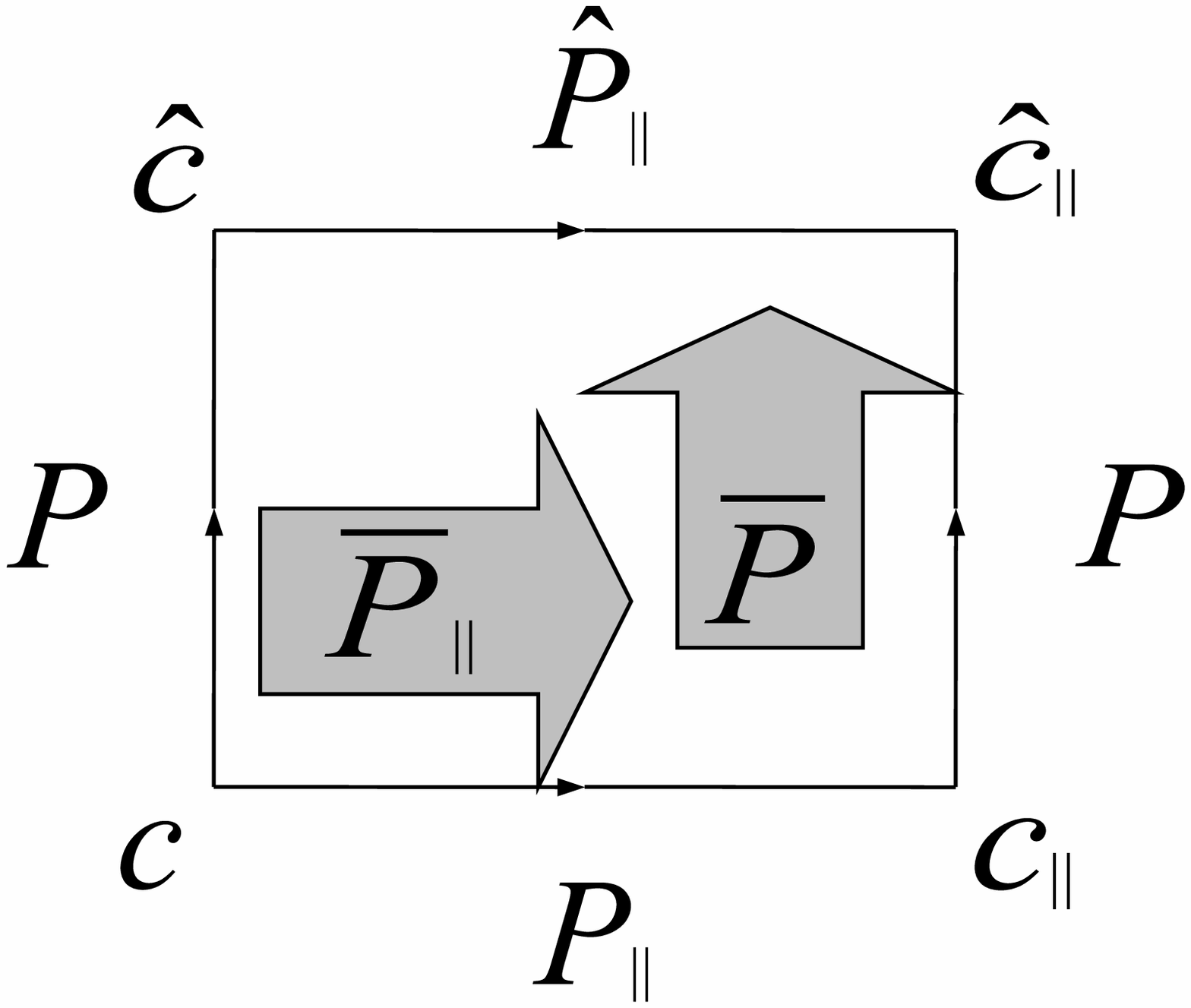}\vspace{-.4in}\\
  \end{center}\caption{{\small Schematic representation of the action of operators
  $\mathbf{\overline{P}}$ and $\mathbf{\overline{P}}_\|$. }
  }
  \label{prisms}
\end{figure}

Let us consider four homotopic cell complexes, $(-1)^{|A|}
c^A_{\|}$, $(-1)^{|A|} \hat c^A_{\|}$, $c^A$ and $\hat c^A$, all
representatives of the single cell structure. In particular let $c$
be a complex and $(-1)^{|c|} c_\|$ be a boundary complex which is
homotopic to it via a homotopy $\mathbf{P}_\|$. Let $\mathbf{P}$ be
a homotopy which sends $c$ to $\hat{c}$ and $c_\|$ to $\hat{c}_\|$.
There will also be a homotopy $\mathbf{\hat{P}}_\|$ between
$\hat{c}$ and $(-1)^{|c|} \hat{c}_\|$. Now the cells
$P(\hat{c}^A,c^A)$ and $P((-1)^{|A|} \hat{c}^A_\|,(-1)^{|A|}
c^A_\|)=(-1)^{|A|}\, P( \hat{c}^A_\|,c^A_\|)$ are also homotopic.
Thus there exists a homotopy $\mathbf{\overline{P}}$ between them,
whose restriction acting on $c$'s and $c_\|$'s is $\mathbf{P}$.
Similarly there exists a $\mathbf{\overline{P}}_\|$ sending the
$c$'s to the $c_\|$'s. The maps are depicted schematically for
0-cells in figure \ref{prisms}.
This diagram shows that the relation
$\mathbf{\overline{P}}_{\|}\mathbf{P}=\mathbf{\overline{P}}\mathbf{P}_{\|}$,
of composition of maps, should hold. This relation is useful in
detailed calculations.

Consider now a homotopy of the $D$-chain $W$ in (\ref{full_W}), that
is, as we have defined it, a homotopy of its dual cell-valued
coefficients. Let that be effected by an operator
$\mathbf{\overline{P}}$ as defined above. Then we have that
\begin{equation}\label{homotopy_full_W}
\partial
\mathbf{\overline{P}}W=\hat{W}-W-\mathbf{\overline{P}}(\partial
W)\,.
\end{equation}
Formula (\ref{bdy of W}) tells that the last term on r.h.s. equals
to
\begin{equation}\label{}
-\sum_A (\mathbf{\overline{P}} c_{\|}^A) \times X_A \,.
\end{equation}
that is the dual cell coefficients are prisms $P(\hat
c_{\|}^A,c_{\|}^A)$.

Now require that the boundary complexes be one and the same: $\hat
c_{\|}^A=c_{\|}^A$. Then their prism $P(\hat
c_{\|}^A,c_{\|}^A)=\varnothing$. Thus we have:
\begin{theorem}
Let $\{\Sigma_A\}$ be a cell complex with $M=\bigcup_A \Sigma_A$ a
manifold with a boundary $\partial M=\bigcup_A X_A$. Fix a dual cell
complex $c_\|$ for the boundary cells $X_A$. Define $D$-chains on
the direct sum of the $\Sigma$ cell complex (with dual cell-valued
coefficients $c$) and $X$ cell complex (with coefficients
$P(c_{\|},(-1)^{|c|}c)=(-1)^{|c|}\mathbf{P}_{\|}c$). Their homology
class is a homotopy invariant, in the way of the coefficients, under
all homotopies that leave $c_\|$ unchanged.
\end{theorem}
In the spirit of the definition \ref{secondary manifold class}, we
define here in the same way but more generally, the secondary
manifold class $\mathcal{W}$ to be this homology class. The
secondary manifold is any representative of this class. In the next
section we give an interpretation of those different forms of the
secondary manifold.

We close the section by discussing an important use of the results
obtained here, and simultaneously take care of a difficulty which we
face when considering real examples.

In order to formulate and prove the well defined smooth limit of
non-smooth configurations, we integrate Lagrangians over the whole
of $M$ and $W$ as a necessary ingredient in the derivations. That is
not satisfying because (i) for $M$ with or without boundary, it
brings in the asymptotic behavior of the fields which might be such
that the action diverges and (ii) it makes completely local
questions like the smoothability of a local discontinuity dependent
on what happens on the whole of $M$.

Being able to treat a manifold with boundary it means that we can
integrate Lagrangians over arbitrary regions $R \subset M$ with
smooth boundary $\partial R$ such that no singularities unrelated to
the discontinuities themselves are involved. That of course makes
our considerations as local as we want.

\subsection{`Renormalization'}\label{Renormalization}

In simple examples the non-uniqueness of $W$ arises in the form of
multiple but apparently equivalent choices for the dual cells as
chains of simplices.

\begin{figure}[h]
  \includegraphics[width=0.45\textwidth, angle=0]{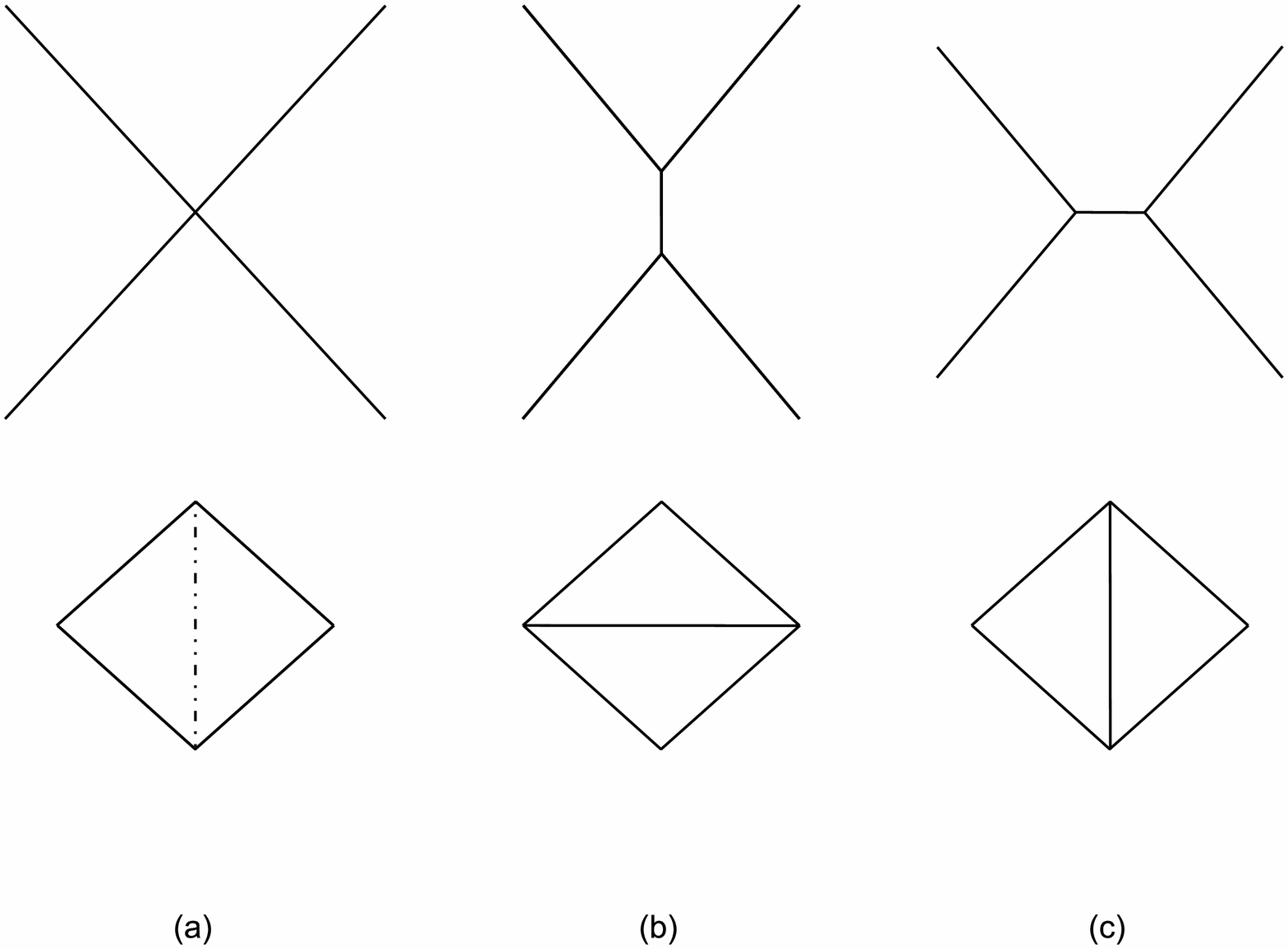}\\
  \caption{{\small  }
  }\label{4way}
\end{figure}

Such an example is shown at the top of figure \ref{4way}a, a cross
section of a 4-way intersection of hypersurfaces. Its dual cell is
shown at the bottom (in solid line). That cell can be expressed as a
chain of simplices in at least two ways: one may divide it along any
of its diagonals. In the figure we draw one of these two cases. The
diagonal is \emph{not} dual to a hypersurface and it is shown by a
dashed line.

These two abstract chains can be thought of as chains residing in
the geometric cell complexes shown at the bottom of figures
\ref{4way}b and \ref{4way}c. In each case they are the sum of the
two 2-simplices in the cell complex. Any one of the two chains can
be used in writing down explicitly the secondary manifold.

\begin{figure}[h]
  \includegraphics[width=.4\textwidth, angle=0]{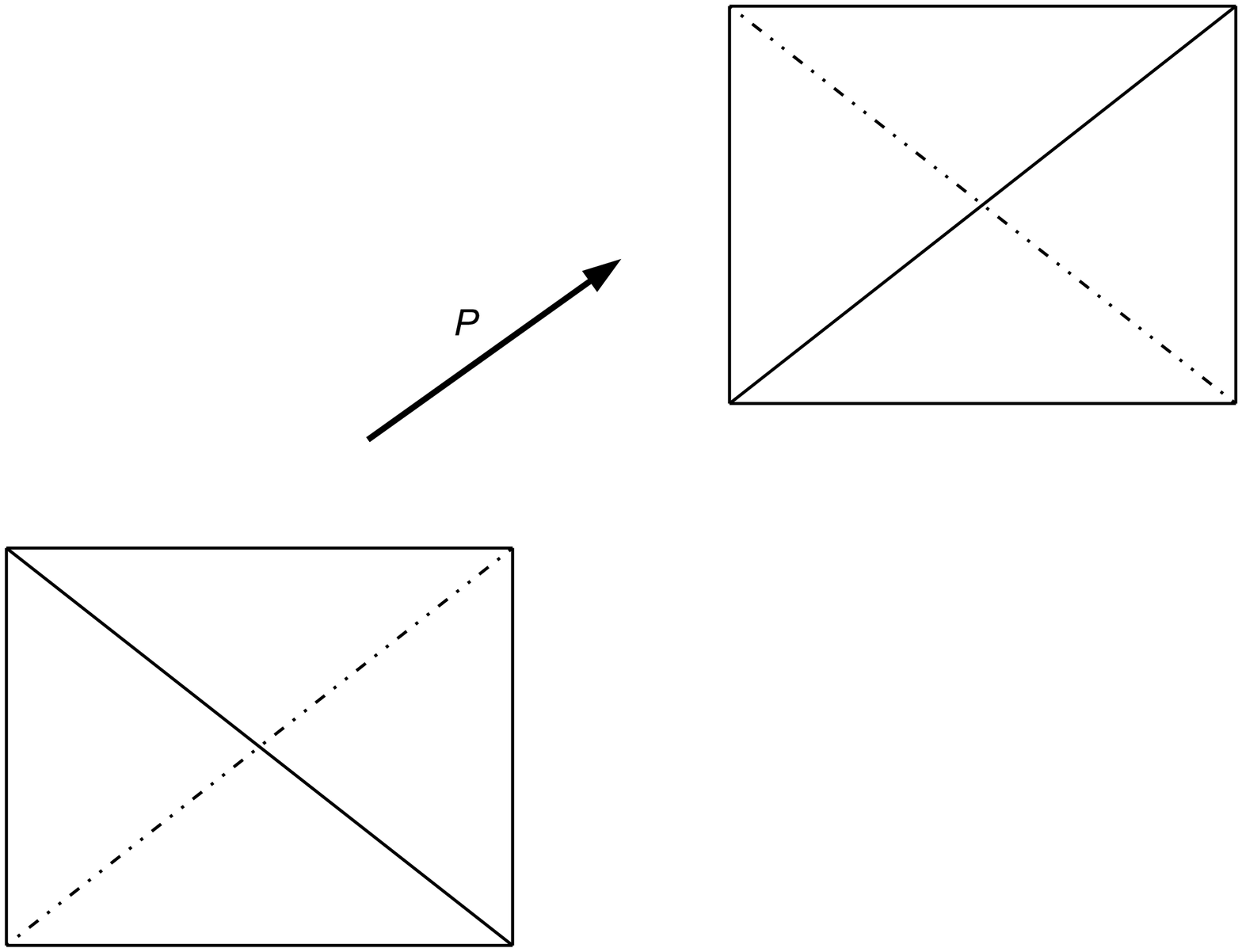}\\
  \caption{{\small  }
  }\label{4wayhomotopy}
\end{figure}

These chains are homotopic. One constructs a prism `interpolating'
between the corresponding abstract cells. In our case this is a
singular parallelepiped (that is, some general deformation of the
Euclidean one). A prism is the value of a homotopy map evaluated at
a cell. So there is a homotopy between the cells. Each cell is a
different chain of simplices. I.e. it is triangulated. We may
further subdivide each cell to reach a \emph{common} triangulation
between them. This is shown in figure \ref{4wayhomotopy}.
So we can `match' their simplices. It is then a mere technicality to
explicitly construct a chain-homotopy between them, which is
standard in the literature~\cite{Hatcher}. Thus one proves the
homotopy of cells as chains. One can construct the simplices we
started with from the smaller simplices of the finer triangulation.

Thus the difference in the choices we have when writing down the
secondary manifold in the 4-way intersection is about homotopy of
the dual chains. As we saw in the previous sections this means that
the two secondary manifolds differ by a boundary. We also saw that
starting from the latter point it is natural to think in terms of
the whole homotopy class of the dual chains, that is, the dual cell
structure~\footnote{To think concretely the homotopies $\mathbf{P}$
that essentially concern us are homotopies of homeomorphisms between
the otherwise abstract cells to chains residing on the large
geometric simplex $S$.}.

Up to this point the `Compton scattering' configurations in the
figures \ref{4way}b and \ref{4way}c have been neglected. They are
there because they turned out to have dual cell complexes homotopic
to that of the 4-way intersection(i.e. also to each other). The
picture is suggestive as to why this is so: These intersection can
be deformed, by continuously `collapsing' the intermediate
hypersurface, to become the 4-way intersection. We can also imagine
the reverse course of deformation, or them deforming to each other.
(I.e. they are homotopic themselves.) The more complicated `Compton'
configurations, seen from a `distance', they will look just like the
4-way intersection. The other way around, from a distance we might
view the latter as two connected more elementary (simplicial)
intersections with vanishing separation.

This suggests something: It might be the case that the many
different homotopic dual cell complexes possible for a given cell
complex $\{\Sigma\}$ are remnants of the many cell complexes that,
when viewed from a distance, look like $\{\Sigma\}$. Or, put
differently, those that become $\{\Sigma\}$ in the limit of
appropriate deformations.

In the rest of this section we shall show that the general effect of
such limits is to move the secondary manifold around within its
class $\mathcal{W}_\Sigma$. This will provide the interpretation we
will looking for in question \ref{secondary_class_question}. Before
doing so we describe another kind of characteristic example.

\begin{figure}[h]
  \vspace{-.0in}\begin{center}
  \includegraphics[width=.49\textwidth, angle=0]{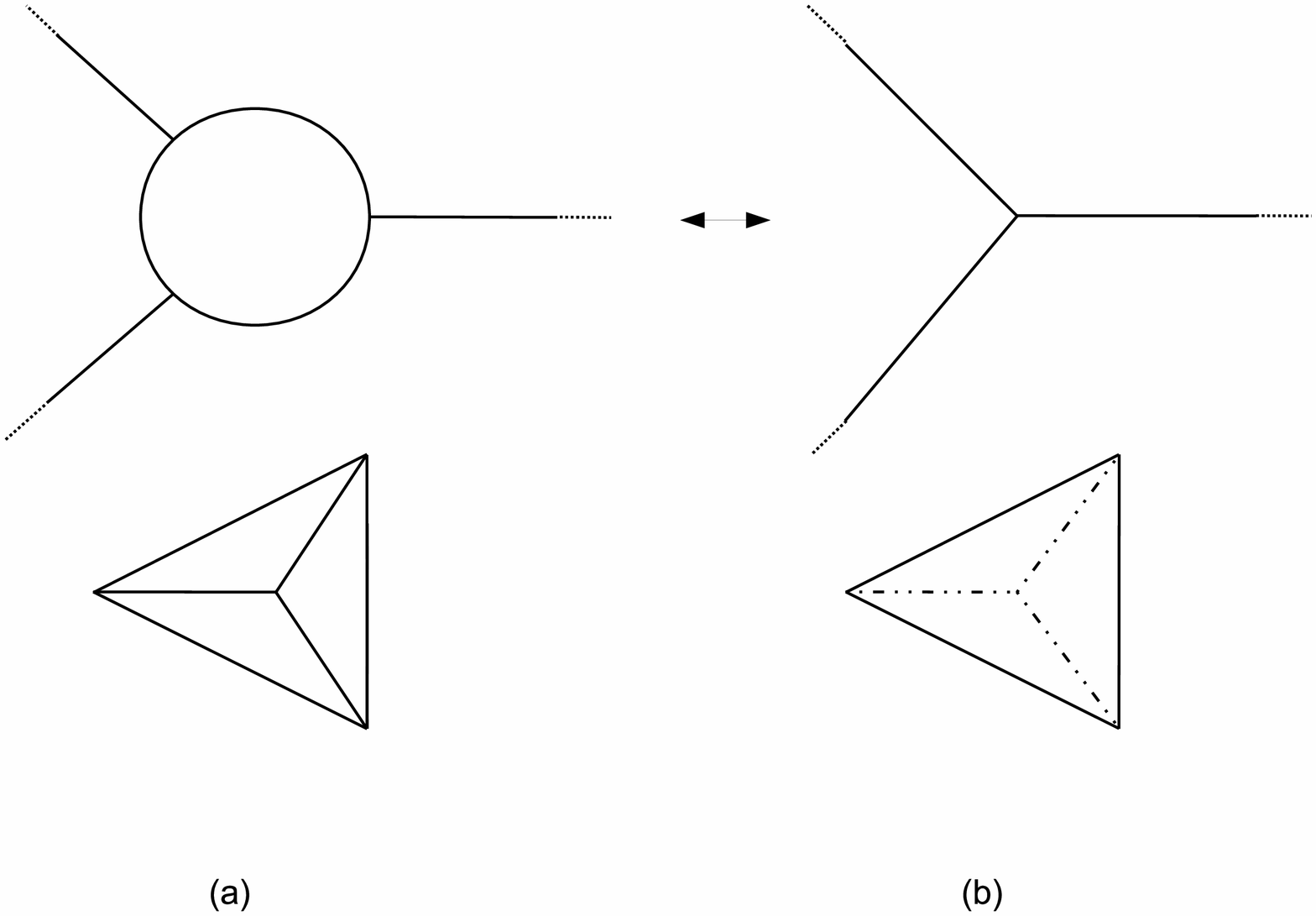}\\
  \end{center}\vspace{-.4in}
  \caption{{\small  }
  }\label{loop}
\end{figure}

The configurations in figure \ref{4way} can be described as `tree
level' deformations of each other. Instead, consider the
configuration at the top of the figure \ref{loop}a. Collapsing the
central loop to zero size the configuration deforms to the 3-way
intersection on the right. Its dual cell complex (in solid line) is
homotopic to the cell complex of the loop configuration: subdividing
it up as the dashed lines show in figure \ref{loop}b we can `match'
the smaller simplices.
These deformations can be described as `loop level' ones and
introduce a new feature: the subdivision of the dual cell complex
introduces a new vertex. On the cell complex (configurations) side
and in the reverse course, a bulk region is collapsed. Seen from a
distance, a loop with vanishing size could be taken for the 3-way
intersection. Thus we may consider a limit that does that. The two
dual complexes being homotopic, we will end up with a secondary
manifold which differs only by a boundary from the old one, i.e. we
deal with two elements of the same secondary manifold class.

At this point, we might wonder whether such `nearby' configurations
translate necessarily to nearby descriptions at the level of fields
satisfying their equations. This matter will be become rather
straightforward after we have established what happens at the level
of the secondary manifold, to which we proceed.

Let $\{\bar \Sigma_I\}$ be a cell complex such that $\bigcup_I \bar
\Sigma=M$ point set-wise. Let a continuous map $\mu : M \to M$ which
shifts point in a continuous manner such that now $M=\bigcup_A
\Sigma_A$ for a cell complex $\{\Sigma_A\}$. I.e. $\mu$ is a
continuous deformation of the net of cells $\bar\Sigma_I$. Some of
them will be driven to extinction, at least as cells of co-dimension
$|I|$.

In general $\mu$ will act as a deformation retraction for
collections of cells $\bar\Sigma_I$, of highest co-dimension $|A|$,
for some $A$, to a cell $\Sigma_A$. By construction, this happens
for all $A$. These collections of cells form sub-complexes of
$\{\bar\Sigma_I\}$ which we denote by $\{\bar\Sigma\}_{\!A}$. We
defined them to be the largest ones with the required property. Thus
they cover the initial cell complex: $\bigcup_A \{\bar
\Sigma\}_{\!A}=\{\bar \Sigma_I\}$ as sets of cells. Note that
$\{\bar\Sigma\}_{\!A}$ are in one-to-one correspondence with
$\Sigma_A$.

In each complex $\{\bar\Sigma\}_{\!A}$ we form chains
$\bar\Sigma_A:=\sum_I \bar\Sigma_I\, \mu(I,A)$, of co-dimension
equal to $|A|$ i.e. built out of the highest co-dimension cells in
the complex. That is, the numbers $\mu(I,A)=0$ unless $|I|=|A|$.
Moreover, we require them to satisfy
\begin{equation}\label{incidence_mu}
\sum_I \bar\epsilon(J,I)\, \mu(I,A)=\sum_B \mu(J,B)\,
\epsilon(B,A)\,,
\end{equation}
where $\epsilon$ and $\bar\epsilon$ are the incidence number
matrices of the cell complexes $\{\Sigma_A\}$ and $\{\bar\Sigma_I\}$
respectively. In matrix notation (\ref{incidence_mu}) reads
$\bar\epsilon \mu=\mu \epsilon$.

The idea is of course that in each complex $\{\bar\Sigma\}_{\!A}$
only the cells appearing in $\bar\Sigma_A$ will not extinct under
$\mu$, or `collapse' as we shall prefer to say. They will simply be
superposed to form $\Sigma_A$. By $\mu(I,A)$ we fiddle the
orientations of this superposition.

We shall call such a complex $\{\bar\Sigma_I\}$ a
\emph{renormalization} of the cell complex $\{\Sigma_A\}$.

By definition chains over a cell complex form a vector space. We
define \emph{a} linear map from $\{\bar\Sigma_I\}$ to
$\{\Sigma_A\}$, meaning between the respective spaces of chains,
denoted also by $\mu$, which encodes the effect of the retraction
$\mu$ on points but also assumes linearity acting on cells.
Specifically we define it such that $\mu(\bar\Sigma_A)=\Sigma_A$,
though this just a matter of convenience. More importantly, for
cells $\bar\Sigma_I$ that extinct under the retraction, at least as
cells of the specific co-dimension, we set\footnote{This shall
always mean that they extinct as cells of co-dimension $|I|$.}
$\mu(\bar \Sigma_I)=\varnothing$.

\begin{corollary}\label{chain map}
The linear map $\mu : \{\bar\Sigma_I\} \to \{\Sigma_A\}$ which
encodes the effects of the retraction $\mu$ is a chain map i.e.
$\partial \, \mu=\mu \, \partial$.
\end{corollary}

\textit{Proof}. Let a cell $\{\bar\Sigma_I\}$. Under `collapse'
three things may happen: i) The cell itself collapses, ii) a cell
$\bar\Sigma_J$ in its boundary $\partial \bar\Sigma_I=\sum_J\,
\bar\Sigma_J\,\bar\epsilon(J,I)$ collapses, iii) i) and ii) do not
apply even inductively.

In iii) $\mu$ acts as the identity so it definitely commutes with
$\partial$. In i) $\partial \mu (\bar\Sigma_I)=\partial
\varnothing=\varnothing$. Also $\mu(\partial \bar\Sigma_I)=\sum_J
\mu(\bar\Sigma_J)\,\bar\epsilon(J,I)$. $\bar\Sigma_I$ collapses thus
$\bar\Sigma_J$ either collapse themselves, or retract to a single
highest co-dimension cell. In the latter case $\mu(\bar\Sigma_J)$
will cancel each other due to opposite orientations. In both cases
$\mu(\partial \bar\Sigma_I)=\varnothing=\partial \mu
(\bar\Sigma_I)$.

In ii) $\mu(\partial\bar\Sigma_I)=\sum_{J'}\,
\mu(\bar\Sigma_{J'})\,\bar\epsilon(J',I)$, where $J'$ sum over all
relevant $J$ except those that $\mu(\bar\Sigma_J)=\varnothing$. On
the other hand the boundary $\partial$ of $\mu(\bar\Sigma_I)$
includes those cells of one co-dimension higher except those that
$\mu(\bar\Sigma_J)=\varnothing$. That is $\partial
\mu(\bar\Sigma_I)=\sum_{J'}\,
\mu(\bar\Sigma_{J'})\,\bar\epsilon(J',I)=\mu(\partial\bar\Sigma_I)$.
Thus in all cases we have that $\partial \, \mu=\mu \,
\partial$.~$\square$

The coefficients $\mu(I,A)$ in the definitions of the chains
$\bar\Sigma_A$ are required to satisfy the relations
(\ref{incidence_mu}) so that the chains satisfy $\partial
\bar\Sigma_A=\sum_B\, \bar\Sigma_B\, \epsilon(B,A)$. Then corollary
\ref{chain map} allows us to deduce that $\partial
\mu(\bar\Sigma_A)=\sum_B\, \mu(\bar\Sigma_B)\, \epsilon(B,A)$. Thus
$\mu(\bar\Sigma_A)=\Sigma_A$ is a consistent requirement on $\mu$.
These two requirements are adequate conditions to translate the
correspondence between $\{\bar\Sigma\}_{\!A}$ and $\Sigma_A$ from
the level of sets to the level of cell complexes.

Let cells $c^A$ and $\bar c^I$ dual to the cells $\Sigma_A$ and
$\bar\Sigma_I$ respectively. Let also cells $\bar{c}^A$ dual to the
chains $\bar\Sigma_A$. The relations (\ref{incidence_mu}) allows us
to show that the chains $\sum_A\mu(I,A) \bar c^A$ have the same
boundary rules as the cells $\bar c^I$. According to our assumptions
the dual structures are unique. That is, those two cell complexes
must be homotopic. The same apply between the cell complexes $\bar
c^A$ and $c^A$. Thus \emph{a} $\bar c^I$ should be equal to
$\sum_A\mu(I,A) c^A$ up to homotopies of the cells.

Note also the following. In matrix notation (\ref{incidence_mu})
reads $\bar\epsilon \mu=\mu \epsilon$. Similarly (\ref{indidence
numbers constraint}) may be written $\epsilon \epsilon=0$ and $\bar
\epsilon \bar \epsilon=0$. Thus any given matrix $\mu$ satisfying
(\ref{incidence_mu}) is one member of an equivalence class of such
matrices:  $\{\mu \thicksim \tilde\mu=\mu+\bar \epsilon \Pi+\Pi
\epsilon\}$ where $\Pi(I,A)$ is any matrix of the right dimension.
That is, any chain $\sum_A \tilde\mu(I,A) \bar c^A$ has the same
boundary rules as the cells $\bar c^I$. One may verify that
directly.

Thus in general we may write
\begin{align}\label{cica}
\bar c^I  = & \sum_A\,   \Big\{\tilde\mu(I,A) \big(c^A+\partial
d^A+\sum_B (-1)^{|B|+1}\epsilon(A,B) d^B\big)\Big\} \nonumber
\\ & - \partial
\bar d^I-\sum_J\,  (-1)^{|J|+1} \bar\epsilon(I,J) \bar d^J\,.
\end{align}
$d^A$ is a cell of dimension $|A|+1$. It can be thought of as the
image of $c^A$ under a chain-homotopy, $d^A=\mathbf{P}c^A$.
Similarly for $\bar d^I$. Note that $\bar d^I$ are general cells
homeomorphic to chains of dimension $|A|+1$ of the geometric simplex
$S$, no different than $d^A$.

Now, are \emph{all} cells of the complex $\{\bar c^I\}$ represented
this way?

The answer is \emph{no}. In detail what happens is: For cells of the
complex $\{\bar \Sigma_I\}$ which extinct under collapsing to
$\{\Sigma_A\}$, there will not be a way to manufacture their duals
from the dual cells of $\{\Sigma_A\}$. We have two cases. First,
those $\bar c^I$ which are not dual to collapsed bulk regions i.e.
they are not vertices. They can be effected by prisms $\bar d^I$
which are \emph{cones}. Secondly, those vertices which are dual to
collapsed bulk regions. Let us denote them by $\bar\Sigma_{\bar i}$
and the dual vertices by $\langle \bar i\rangle$.

Having picked a dual complex $\bar c^I$ to the complex $\bar
\Sigma_I$ we can write down the secondary manifold: $\bar W=\sum_I
\bar c^I\times\bar \Sigma_I$. Using (\ref{cica}) and bearing in mind
the previous remarks one finds:
\begin{equation}\label{bar W}
\bar{W}=\sum_A c^A \times \bar \Sigma_A+ \partial \bar Y+\sum_{\bar
i} \langle \bar i \rangle \times \bar \Sigma_{\bar i}\,,
\end{equation}
where the manifold $\bar Y$ is explicitly given by
\begin{align}\label{}
& \bar Y=\sum_I  f^I  \times \bar \Sigma_I\,, \\
&  f^I=  \sum_B\big\{(-1)^{|B|}\,\Pi(I,B)\, c^B+\tilde \mu(I,B)\,
d^B\big\}+\bar d^I\,. \nonumber
\end{align}
To derive this result we use the fact that $\epsilon(A,B)=0$ unless
$|A|=|B|+1$, which allows us to write the boundary rules in the form
$\sum_B\epsilon(A,B)c^B=(-1)^{|A|}\partial c^A$, and the properties
of the coefficients $\mu(I,A)$.

Define a natural extension of $\mu$ acting on chains of
$\{\bar\Sigma_I\}$ from number-valued to dual cell-valued
coefficients to be the linear map, denoted again by $\mu$, defined
by
\begin{equation}\label{}
\mu(\bar c^I \times \bar\Sigma_J):=\bar c^I \times
\mu(\bar\Sigma_J)\,.
\end{equation}
One may verify that this is a chain map: $\partial\, \mu=\mu\,
\partial$.

$\bar \Sigma_{\bar i}$ are the collapsed bulk regions. That is
$\mu(\bar \Sigma_{\bar i})=\varnothing$. We have
\begin{equation}\label{final_collapse}
\mu(\bar W) = \sum_A c^A \times \Sigma_A +
\partial  Y\,,
\end{equation}
where we defined $Y=\mu(\bar Y)$. $Y$ is a chain of dimension $D+1$
over the complex $\{\Sigma_A\}$ with cell-valued coefficients. The
chain $\sum_A c^A \times \Sigma_A$ is a secondary manifold for this
complex.

In more general terms our results can be stated as follows.
\begin{theorem}\label{collapsed_renormalization_theorem}
Let $\{\Sigma_A\}$ be a cell complex with $M=\bigcup_A \Sigma_A$ a
manifold without boundary, and $\mathcal{W}_{\Sigma}$ its secondary
manifold class. Let a renormalization $\{\bar \Sigma_I\}$ of
$\{\Sigma_A\}$ with a secondary manifold class
${\mathcal{W}}_{\bar\Sigma}$. Then
$\mu(\mathcal{W}_{\bar\Sigma})=\mathcal{W}_\Sigma$.
\end{theorem}

The result (\ref{final_collapse}) provides an answer to the question
\ref{secondary_class_question}. We have a general process producing
arbitrary cobordisms $Y$ between elements of the secondary manifold
class $\mathcal{W}$. Thus we may interpret the different secondary
manifolds in $\mathcal{W}$ as different collapsed renormalizations
of any given secondary manifold $W$ in $\mathcal{W}$.

\subsection{Renormalization and action}\label{Renormalization and
action}

Let now ${\cal K}$ be the secondary Lagrangian of a certain field
theory defined on a cell complex $\{\Sigma\}$. The secondary
Lagrangian ${\cal K}$ can be thought of as a linear map from the
union of all cells $\{c \times \Sigma\}$ to the real numbers. The
value of this map acting on a chain $W$ belonging into a class
${\cal W}_\Sigma$ is the secondary action. For any two cell
complexes $\Sigma$ and $\bar \Sigma$ the classes
$\mathcal{W}_\Sigma$ and $\mathcal{W}_{\bar \Sigma}$ are in general
unrelated. Therefore so are the secondary actions.

We investigate the behavior of the theory under collapsing. It is
helpful to give collapsing a sense of progression. We invent a
sequence of complexes $\bar \Sigma^\lambda_I$ which are different
`instants' of the collapsing effected by the deformation $\mu$. That
is $\{\bar \Sigma_I^\lambda\} \to \{\Sigma_A\}$ as $\lambda \to
\infty$. In particular $\lim_{\lambda \to \infty} \bar
\Sigma^\lambda_I=\mu(\bar\Sigma_I)$. It is adequate to denote the
sequence also by $\mu$. We investigate the limit of the map ${\cal
K}$ over such sequences.

Define the secondary manifolds $\bar W^\lambda=\sum_I \bar c^I
\times \bar\Sigma_I^\lambda$. Collapsing the cells $\bar\Sigma_I$ we
end up with ${\Sigma_A}$. We pick a complex $c^A$ dual to
$\Sigma_A$. The cells  $c^A$ must be related to $\bar c^I$ by a
relation (\ref{cica}).

We may re-write $\bar W^\lambda$ in the form (\ref{bar W}), with
$\bar Y^\lambda=\sum_I f^I \times \bar \Sigma^\lambda_I$. The effect
of retraction is given by a relation (\ref{final_collapse}),
$\mu(\bar W^\lambda)=W+\partial Y$. $Y$ depends only on $\mu$ i.e.
the sequence, not on the `instants' labeled by $\lambda$ .

The manifold $Y$ might not be unique in general: the cells $d^B$ and
$\bar d^I$ give a lot of freedom in the way one writes the relation
(\ref{cica}) between the chosen complexes $\bar c^I$ and $c^A$.
Moreover, $Y$ definitely depends on the sequence of cells $\bar
\Sigma$. An example of this elementary fact initiated our discussion
in section \ref{Renormalization}, figure \ref{4way}.

Now consider the limit
\begin{equation*}
\lim_{\lambda \to \infty}\int_{\bar W^\lambda} {\cal K}\,,
\end{equation*}
for an arbitrary collection of sequences $\mu$. Applying equations
(\ref{bar W}) and (\ref{final_collapse}) to $\bar W^\lambda$ we have
that this limit equals
\begin{equation}\label{xxx}
\int_W {\cal K}+\int_Y (\delta+d){\cal K}+\lim_{\lambda \to \infty}
\sum_{\bar i}\int_{\bar \Sigma^\lambda_{\bar i}}{\cal L}\,.
\end{equation}
$Y$ depends on the sequence $\mu$, thus the limit is independent of
the sequence if
\begin{equation}\label{non-renormalizable}
\int_Y (\delta+d){\cal K}=0\,.
\end{equation}
This holds modulo the last term in (\ref{xxx}) which does not exist
at the level of secondary manifold relations and requires some
attention.

The statement $\lim_{\lambda \to \infty} \bar\Sigma^\lambda_{\bar
i}=\mu(\bar\Sigma^\lambda_{\bar i})=\varnothing$ need not be
respected by the integrals where we integrate functionals of fields.
The very essence of such a statement is that each bulk cell is
\emph{contractible}, i.e. everywhere in the interior of
$\bar\Sigma^\lambda_{\bar i}$ the fields $\Psi_i$ are smooth and no
singularities arise. If they did, we would be forced to exclude
points from those cells. Then work differently to find if possible
the right Lagrangian terms associated with such singularities. That
would be the case if the cells contained conical singularities.
Explicitly, we shall assume that our fields belong to the field
space $\Phi_\Sigma$, as defined below. Then Lagrangians ${\cal L}$
are well behaved and the last term in (\ref{xxx}) vanishes.

The fields are arbitrary thus condition (\ref{non-renormalizable})
translates to local statements at each cell $\Sigma_A$.
\begin{theorem}[Non-renormalization theorem]\label{renormalization theorem}
Let $\{\Sigma\}$ be a cell complex and a secondary manifold $W$. The
secondary action $\int_W {\cal K}$ is the well defined limit of
renormalizations of $\{\Sigma\}$ retracting to it iff
\begin{equation}\label{non-renorm}
i^*_{\Sigma_A}\left(\int_{d^A}(\delta+d){\cal K}\right)=0
\end{equation}
for all $\Sigma_A \in \{\Sigma\}$.
\end{theorem}

The cells $d^A$ are linear combinations of $f^I$ defined through
$\sum_I f^I \times \mu(\bar\Sigma_I)=\sum_A d^A \times \Sigma_A$.
What matters of course is that they are some cells of dimension
$|A|+1$.

The prisms $d^A$ make any cell $c^A$ homotopic to cells with an
arbitrary number of additional vertices. In the interpretation of
collapsed renormalizations these are specifically thought of as dual
remnants of collapsed bulk regions. ${\cal K}$ involves explicitly
the fields associated with them.

Condition (\ref{non-renorm}) cannot be satisfied without certain
requirements on those `ficticious' fields. Their very interpretation
as fields of bulk regions which shrank to zero volume makes the
following a natural choice: We require that all fields appearing in
the secondary Lagrangian satisfy the same kind of conditions. Thus
we manage to put condition (\ref{non-renorm}) in some sense on the
same basis as the smoothing criterion \ref{smoothing theorem}.

The previous statement can be phrased more carefully. Let's at this
point be more specific about our general conditions on the fields.

Given a complex $\{\Sigma_A\}$ covering the manifold $M$ and a
collection of fields $\Psi$ over $M$, by \emph{field space}
$\Phi_\Sigma$ we shall mean the set of all configurations of $\Psi$
such that they are at least $C^r$ in the interiors of $\Sigma_i$ and
$C^{1-}$ i.e. have at most finite discontinuities across the
boundaries of $\Sigma_i$. The number $r$ is the maximum order of
derivatives of that field in the Lagrangian.

Given a Lagrangian (theory) ${\cal L}(\Psi)$ defined over $M$, by
\emph{smoothable field space} $V_\Sigma$ associated with ${\cal
L}(\Psi)$ we mean a subset of the field space where the secondary
Lagrangian ${\cal K}(\bar\Psi)$ of the theory is smoothable. (That
implies that all fields in that subset are smoothable field
configurations of the theory.\footnote{The maximal $V_\Sigma$ of
vacuum solutions, or better the union of $V_\Sigma$ for all
complexes $\Sigma$, is the natural field space over which we should
integrate in a path integral in any sensible theory with covariant,
discontinuous vacuum solutions. Clearly such theories must be more
or less topological.}).

Applying the smoothing theorem we require that the field
$\Psi_\alpha$, which approximates the discontinuous configurations
of the fields $\Psi_i$, is arbitrary. This guaranties that our
results are independent of how one approximates the configuration
$\{\Psi_i\}$. This is a very strong condition. Working out specific
examples, one finds that there is hardly a way to succeed unless one
imposes continuity on some of the fields, and no conditions on the
rest.

A natural way to do that is by separating the fields in some natural
way: into vielbein and connection, as we did in section
\ref{Lovelock-Cartan Gravity}, or in homolomorphic and
anti-holomorphic components as one may chose to do in Chern-Simons
theory. Roughly, one separates the fields into coordinates and
momenta. Then imposes e.g. continuity on the momenta and nothing on
the coordinates. This implies that a `momentum' $\Psi_\alpha$ must
be continuous and agree\footnote{The limit $\alpha \to \infty$ is
assumed. The smooth limit is guarantied under the vanishing of
quantities which resemble a lot the symplectic form of a given
theory. Then it is not a surprise that a separation of variables
into coordinates and momenta is helpful.} with all momenta $\Psi_i$
at all $\Sigma_A$, while the `coordinate' $\Psi_\alpha$ is
completely free, like all the other coordinates $\Psi_i$. Thus we
have a space $V_\Sigma$ where the field $\Psi_\alpha$ operates in a
completely symmetrical way with the fields $\Psi_i$. Then, in the
dual space, the vertices $\langle\alpha\rangle$ can be treated in a
completely symmetrical way with the other vertices in the complex.

At the level of the dual space the smoothing and non-renormalization
theorems are closely related: they involve cells $d^A$ which are
cones $C(\alpha,c^A)$ and general homotopies $\mathbf{P}c^A$
respectively. Working in field space $V_\Sigma$ such as the ones
introduced in the previous paragraph, one may treat the cells
$C(\alpha,c^A)$ as any of the cells $\mathbf{P}c^A$. That is we may
speak collectively of cells $d^A$. Then the statements of the
smoothing and non-renormalization theorems coincide.

Following our terminology, a theory such that (\ref{non-renorm})
holds we may call it \textit{non-renormalizable}, in the sense that
its description on a cell complex $\{\Sigma\}$ is not different than
the limit of a sequence of renormalizations of $\{\Sigma\}$. We have
shown that, under natural conditions, smoothable is a synonym of
non-renormalizable. This makes sense- to the extent the conditions
are `natural', or rather we use this equivalence to define what we
mean when we say natural: If discontinuous solutions are
legitimately weak limits of continuous solutions of the field
equations, why shouldn't they legitimately deform to each other?
From another perspective, renormalization asks again in a slightly
different way the essential question we posed at the beginning of
our work: Under what conditions does a theory admit distributional
solutions? This is the subject of the next section.

\section{Field equations and general covariance}

\subsection{Smoothable field equations}

The source tensor of our fields, such as stress-energy and spin
tensors, are uniformly defined throughout the manifold $M$ as long
as the Euler-Lagrange variations of the fields are continuous. Thus
the variations, denoted by $\delta_{\mathrm{EL}}\Psi_i$, must be all
equal to a single smooth field, we shall call $\psi$. (Recall that
the secondary action is a functional of the bulk fields $\Psi_i$.)

Also, wanting to compare the field equations of a smooth field
$\Psi_\alpha$ to the secondary field equations of the fields
$\Psi_i$, we impose that
$\delta_{\mathrm{EL}}\Psi_\alpha=\delta_{\mathrm{EL}}\Psi_i$. In all
we require that $\psi:=\delta_{\mathrm{EL}}\bar\Psi$ and
$\delta\psi=0$, where $\bar\Psi$ is defined over the smoothing
manifold $W^\alpha$, or according to the previous section, over any
cobordism $Y$ relating two elements of the secondary manifold class.

The secondary action ${\cal S}=\int_{W} {\cal K}$ being smoothable
means that under certain conditions on the fields
\begin{equation}\label{euler lagrange}
\lim_{\alpha \to \infty}\int_{\langle\alpha\rangle \times M} {\cal
K}=\int_{W} {\cal K}\,.
\end{equation}
Let the Euler-Lagrange variations be done within the space of fields
satisfying those conditions, called $V_\Sigma$ in the previous
section. The conditions defining $V_\Sigma$ constrain the
differences of the fields, and are always such that the fields
$\Psi_\alpha+\psi$ and $\Psi_i+\psi$ also belong to $V_\Sigma$, for
a first order infinitesimal arbitrary smooth field $\psi$. Taking
the difference of the two relations we obtain the convergence of the
Euler-Lagrange variations, or in other words
\begin{equation}\label{euler 2}
\lim_{\alpha \to \infty}\int_{\langle\alpha\rangle \times M} \psi
\cdot \frac{\delta {\cal S}}{\delta \bar \Psi}=\int_W \psi \cdot
\frac{\delta {\cal S}}{\delta \bar \Psi}\,.
\end{equation}
We used an obvious notation for the contraction of the various
indices of the fields $\Psi^{a \cdots}_{\mu \cdots}$. Thus
smoothability of the secondary \emph{field equations} follows.

On the other hand, one may investigate out of curiosity the
alternatively possibility of looking directly at the smoothability
of the secondary field equations. In other words, to analyze the
smooth limit conditions for the functional
\begin{equation}\label{field equations Lagrangian}
\psi \cdot \frac{\delta {\cal S}}{\delta \bar\Psi}\,,
\end{equation}
where $\psi$ is an arbitrary smooth field tensorial in all its
indices. Its \emph{integral} over $W$ coincides with the
Euler-Lagrange variations of the secondary action ${\cal S}$. But we
do not use ${\cal S}$ directly.

One may view the integrand in (\ref{field equations Lagrangian}) as
a new Lagrangian involving a non-propagating smooth field $\psi$.
That is, we treat its integral over $W$ as a new secondary `action'.
To check the smooth limit of field equations themselves, one may
apply the criterion \ref{smoothing theorem} for this `action' for an
arbitrary smooth test field $\psi$. This reads
\begin{equation}\label{`action' smoothability}
\int_{W^\alpha}(\delta+d) \left(\psi \cdot \frac{\delta {\cal
S}}{\delta \bar\Psi}\right) \to 0\,,
\end{equation}
as $\alpha \to \infty$. This is of course nothing but (\ref{euler
2}); we haven't really derived anything new in the last two
paragraphs, only gained a useful point of view.

Since $\delta \psi =0$, the `Lagrangian' (\ref{field equations
Lagrangian}) will contain one less $t$-dependent derivative factor
than the secondary Lagrangian ${\cal K}$, thus it is smoothable
under weaker conditions on the fields.

\subsection{Diffeomorphisms}

Constructing the description of a theory for discontinuous fields
one may have a kind of more basic problem than the description being
not smoothable. The breaking of translational invariance by the
discontinuities could render the whole construction not coordinate
choice independent.

Let $\xi$ be an arbitrary vector field in the tangent space of $M$.
First, the secondary actions are required to be invariant under the
(diffeomorphism) transformations it generates. We found in section
\ref{Finding the secondary Lagrangian} that the invariance of the
action translates to the formula
\begin{equation}\label{}
\int_W i(\xi) (\delta +d) {\cal K}+\int_{\partial W} i(\xi) {\cal
K}=0\,.
\end{equation}
We allow for the possibility that $M$ possesses a boundary, in which
case $W$ is the appropriate secondary manifold constructed in
section \ref{Manifolds with boundary}.

Now let the fields we think of as intrinsic to the boundary be
continuous. That is, the dual complex $c^A_{\|}$ associated to these
fields amounts to a single vertex, $\langle \| \rangle $. Then,
presumably, $\mathbf{P}_{\|}=C(\|,\cdot)$ i.e. in (\ref{full_W}) is
a cone operator. In this case, the second term in the formula above
amounts to an integral over $\partial M$ of $i(\xi) {\cal L}$. That
is, a $D$-form must be built out of fields intrinsic to the boundary
of $M$. Such a form is identically zero.

\begin{theorem}\label{diff theorem}
Let a cell structure $\{\Sigma_A\}$ such that $M=\bigcup_A \Sigma_A$
may possess a boundary. Let a theory with a secondary Lagrangian
${\cal K}$. Let the fields intrinsic to the boundary of $M$ be
continuous, or $\partial M=\varnothing$. Then the secondary action
$\int_W {\cal K}$, with $W$ being given by (\ref{full_W}), is
invariant under diffeomorphism transformation iff
\begin{equation}\label{diff2}
\int_W i(\xi) (\delta +d) {\cal K}=0\,.
\end{equation}
\end{theorem}
As the arbitrary field $\xi$ may have an infinitesimal support we
find that (\ref{diff2}) equivalently reads
\begin{equation}\label{diff3}
\left. \int_{c^A} (\delta +d){\cal K}\right|_{\Sigma_A}=0\,,
\end{equation}
at each cell $\Sigma_A$ integrating over a dual $c^A$.

Thus we have to deal with two field spaces. A subset of the field
space for which the secondary action is smoothable. Also a subset of
the field space where the secondary action is diffeomorphism
invariant. Let their intersection be a not measure zero subset.
(This is the case of the examples we considered in section
\ref{Lovelock-Cartan Gravity}. In the `smoothable field space'
chosen there, we have in general that $(\delta+d){\cal K}=0$ at any
$\Sigma_A$. This is stronger than both required conditions therefore
implies them.) Then (\ref{diff2}) holds for any two configurations
$\Psi$ and $\Psi+\psi$, where $\Psi=(\Psi_\alpha, \Psi_i)$. This
implies
\begin{equation}\label{diff4}
\int_W i(\xi) (\delta +d)\left(\psi \cdot  \frac{\delta{\cal
S}}{\delta \bar \Psi}\right)=0\,.
\end{equation}

This is nothing but the invariance of the `action' (r.h.s. of
(\ref{euler 2})). $\psi$ is an arbitrary smooth field, tensorial in
all its indices. Thus the `action' is invariant iff the field
equations $\frac{\delta{\cal S}}{\delta \bar \Psi}$ are covariant.
Therefore a (smoothable) invariant secondary action implies
(smoothable) invariant secondary field equations.

Secondly, as in the previous section about smoothability, let's not
worry about the secondary action and consider the general covariance
of the field equations themselves. This means: We require
(\ref{diff4}).

This condition is highly restrictive. Very little of the large field
space that `fits' the weak smoothability condition (\ref{`action'
smoothability}) remains after (\ref{diff4}) is imposed.

\subsection{Example: Chern-Simons theory}

Consider as an example Chern-Simons theory: Let a Lie group $G$ be a
gauge group with a connection $A$ on a three-dimensional manifold
$M$ without boundary. Let $N$ be a manifold such that $\partial
N=M$. Define the theory by $\frac{1}{2}\int_N \mathrm{Tr}(FF)$,
where $F=dA+A A$ is the curvature of the connection $A$.
$\mathrm{Tr}$ is an invariant bilinear form of the Lie algebra of
the group. The integrand is an exact form. This integral defines
Chern-Simons gauge theory over $M$. The appropriate choice of
secondary action is such that that $(d+\delta) {\cal K} =
\mathrm{Tr}({\cal F} {\cal F})$ where ${\cal F}=\delta \bar A+\bar
F$. So locally it is given by
\begin{equation*}
 {\cal K} = \mathrm{Tr}\left(\bar{A} (d+\delta) \bar{A} + \frac{2}{3}
 \bar{A}\bar{A}\bar{A}\right)\, .
\end{equation*}
Considering the smoothing of the field equations, condition
(\ref{`action' smoothability}) requires that
\begin{equation}\label{CS smooth field eq}
\int_{W^{\alpha}} \mathrm{Tr}(d\psi {\cal F}) \to 0\,,
\end{equation}
for $\alpha \to \infty$. The integral involves only one factor
${\cal F}$ i.e. a single factor $\delta \bar A$ to be integrated
over the bulk regions $\Sigma_i$. This involves the quantities
$A_\alpha-A_i$ which vanish in the limit. Thus (\ref{CS smooth field
eq}) holds under \emph{no restrictions} on $A_i$. Then the secondary
action itself is not smoothable: That would require vanishing of
$\int_{W^\alpha} \mathrm{Tr}(\delta \bar A\,\delta \bar A)$. Thus we
have well defined secondary equations under conditions in which the
secondary action is not smoothable. This sounds rather strange or
too good to be true, and indeed it is, in the following sense.

If we require that the secondary field equations respect
diffeomorphism invariance, then (\ref{diff4}) implies the strong
condition: $A_i-A_j=0$ at any suface i.e. the field must be simply
continuous. That is, it leaves us with no interesting discontinuous
field space at all. It seems that there is something unnatural in
the idea of admitting discontinuous solutions of field equations for
which the secondary action is not smoothable, as it sends us into a
narrow path. Here in the case of Chern-Simons theory we have seen
that it can be done but the price to pay would be breaking of
diffeomorphism invariance, e.g. by promoting the position of the
surface to a dynamical variable.

Let's write down the conditions explicitly. They take on an
interesting form. Along a surface $\Sigma$ introduce coordinates
$\sigma$ and $\bar\sigma$, and a normal coordinate $\nu$. Thus we
may write $\bar A=\bar A_{\sigma}d\sigma+\bar
A_{\bar\sigma}d\bar{\sigma}+\bar A_{\nu}d\nu$. Smoothability
condition of the action reads
\begin{equation}\label{}
\int_{s}i^*_\Sigma\, \textrm{Tr}(\delta\bar A_\sigma \delta \bar
A_{\bar \sigma})=0\,.
\end{equation}
Diffeomorphism invariance of the action reads
\begin{equation}\label{}
\int_{s}i^*_\Sigma\, \textrm{Tr}(\delta\bar A_\sigma \delta\bar
A_\nu)=0\,, \:\: \textrm{and} \:\:\: \int_{s}i^*_\Sigma\,
\textrm{Tr}(\delta \bar A_{\bar \sigma}\delta\bar A_\nu)=0\,.
\end{equation}
$s$ is a 2-simplex. Both conditions are satisfied if either of the
`canonical conjugates' $A_\sigma$ or $A_{\bar\sigma}$ are
continuous, and $A_\nu$ changes dependent on $A_\sigma$ or
$A_{\bar\sigma}$, or is simply continuous.

\section{Summary}

A singular or non-smooth field can be seen either as an
approximation to a smooth field or as the limiting case of some
family of smooth fields. An important question is whether this
non-smooth configuration captures the relevant details of the smooth
field(s) or whether something is lost. Another, similar, question is
whether a non-smooth field can be regarded as an \emph{exact}
solution of equations of motion. Or we may ask if non-smooth fields
can be admitted in the classical action principle, or what role they
play in the path integral.

Consider different sequences of smooth fields which converge to the
same discontinuous. If for any such sequence, the limiting
contribution to the action is the same, we may say that no
information is lost by identifying the discontinuous field as a
single limit point in the space of fields. If, on the contrary,
different sequences give different limiting values of the action,
then there is some microstructure that is being missed and which is
contributing to the action. Take the example of a shell. One can
imagine constructing various thick shells with different internal
profiles for the field, which all converge to some step function.
Now consider the limit in which the thickness of the shell goes to
zero. Do the details of the internal profile continue to affect the
value of the action in the limit?

We have discussed the problem in terms of the classical action
principle or the path integral (assuming that it can be defined).
The Lagrangian is a function of the field and it is this integral
which we are concerned with. The question of whether a function of
the field can be multiplied by a smooth test field and integrated is
not directly relevant since the test field does not exist in the
physical problem.

If the different limiting processes give different contributions,
then it can be regarded as an approximation to the solution, but we
are not able to say which physically distinct limit process we have
taken. So it is hard to think of it as an exact solution. So it is
not clear whether one should include such fields in the classical
action principle. In the path integral, it is a different matter:
the question seems to have to do with how many times the field
configuration is counted and whether the contributions ``wash out"
or whether they can add up if the phase is approximately stationary.

These questions are conveniently addressed in a geometrical setting.
The manifold $M$ is a place where the field $\Psi$ is discontinuous.
Discontinuities divide $M$ up into a network of cells of various
dimensions, a cell complex. On each bulk cell $\Psi$ is continuous
with a value $\Psi_i$. A space $\sum (\textrm{dual cell}) \times
(\textrm{cell})$ is a place where a new field $\bar\Psi$ is
continuous. $\bar \Psi$ changes continuously along the dual cells
relating the $\Psi_i$'s across the discontinuities. It would be
difficult to proceed if the inverse of $\Psi$ is involved in the
Lagrangian ${\cal L}(\Psi)$. Suppose that it does not. Then a
Lagrangian for $\bar \Psi$ can be constructed: it is given by the
same rule ${\cal L}: \Psi \mapsto {\cal L}(\Psi)$, only replacing $d
\to d+\delta$, the derivative operator on the new space. This new
Lagrangian, ${\cal K}$, the secondary Lagrangian, can give an action
for the fields $\Psi$ if integrated over $W\equiv \sum (\textrm{dual
cell}) \times (\textrm{cell})$, the secondary manifold.

\begin{figure}[h]
  \includegraphics[height=0.30\textwidth, angle=0]{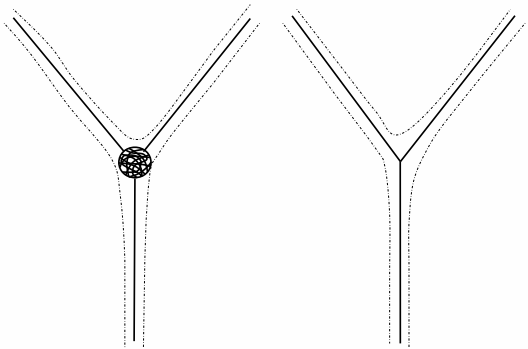}\\
  \caption{{\small Fixing a length scale, the length between the dashed lines in
  the sketch above, two different configurations may look the same.
  A \textit{smoothable} theory, or `non-renormalizable', is such that the respective
   solutions converge to a single solution
  when this scale goes to zero.
 This can be read in reverse: fixing a scale, we may replace a certain
 configuration by another whose intersections are all of the most basic type, i.e. whose dual
 cells are simplices. An example was given by figure \ref{4way}. Configuration
 \ref{4way}a can be viewed as a combination of the ones in \ref{4way}b and
 \ref{4way}c. The error in the two solutions treated as the same thing is
smaller the smaller that scale is.
 }
  }\label{renorma}
\end{figure}
The central property of the secondary manifold is: if $\partial
M=\varnothing$ then $\partial W=\varnothing$. Given a division of
$M$ into cells, a fixed cell complex, property $\partial
W=\varnothing$ is defining $W$. That is, $W$ is a representative of
a homology class. $M$ belongs to this class, written appropriately.
Given a continuous field $\Psi_\alpha$ over $M$ we may associate a
dual vertex $\langle\alpha\rangle$ with it and write $M$
isomorphically as $\sum\: \langle\alpha\rangle \times (\textrm{bulk
cell})$. This is indeed a $W$. Integrating ${\cal K}$ over this
space we get nothing but the usual action evaluated for
$\Psi_\alpha$. There is a cobordism between this space and any $W$,
the smoothing manifold $W^\alpha$. Imagine $\Psi_\alpha$
approximating better and better  a given discontinuous configuration
$\Psi_i$ as $\alpha \to \infty$. Recall that the fields $\Psi_i$
were the starting point for introducing the secondary action $\int_W
{\cal K}$. Stokes' theorem allows us to relate it with the usual
action for $\Psi_\alpha$: they converge to each other iff
$\int_{W^\alpha}(d+\delta){\cal K}$ vanishes as $\alpha \to \infty$.
This is the smoothing theorem \ref{smoothing theorem}. Additionally,
there are many other $W$'s in the homology class and many cobordisms
$Y$ between them. The different $W$'s can be interpreted as
secondary manifolds-remnants of configurations which have been
collapsed to the given cell complex. Imagine that it happens that
$\int_{Y}(d+\delta){\cal K}=0$ for all cobordisms $Y$. Then the
secondary action is independent of which $W$ it is evaluated on i.e.
which limiting configuration it is evaluated on, as long as it
approaches the fixed cell complex. Configurations of intersecting
hypersurfaces are remeniscent of Feynman diagrams and we can make an
amusing analogy with properties of quantum field theories. Fixing a
length scale, any difference in the value of the action on different
configurations which differ below that scale, will be bounded in the
order of that scale and go to zero with it. The action does not get
`renormalized'. An example is sketched in figure \ref{renorma}.

An independent restriction is imposed by requiring diffeomorphism
invariance of the action. It is a condition again on
$(d+\delta){\cal K}$, theorem \ref{diff theorem}. This is a
condition essentially complementary to smoothability. Cobordisms $Y$
are extending $W$ in the directions of the dual cells, we deal with
dual cells of one dimension higher. Diffeomorphism invariance on the
other hand, due to the inner product entering the equations,
requires dealing with forms of order 1 higher in the space-time
dimensions. Also, for this reason the condition of diffeomorphism
invariance is typically somewhat stronger than smoothing.

Once these conditions are satisfied one is in a position to show
that $\Psi_i$ can indeed be regarded as an exact solution of the
equations of motion of the theory. This is possible as conditions
are to hold `off-shell'. The conditions are typically satisfied if
we separate the fields into canonical coordinates and momenta and
require continuity in either of the canonical variables. This fact
may have some interesting implications in the quantized theory when
it makes sense.

Thus the problem posed can be reduced to a question of conditions
preserving the cohomology of certain forms, the secondary
Lagrangians. I.e. conditions of an essentially topological nature,
though we have concerned ourselves only with purely local questions
in gravitation. The analogy is between the continua of connections
which preserve the Euler number, for example, and the continua of
fields $\Psi_\alpha$ which approximate a given discontinuous
configuration.

\acknowledgments S.W thanks J. Zanelli for many helpful comments.
This work was partially funded by FONDECYT grant 1085323. Centro de
Estudios Cient\'{\i}ficos (CECS) is funded by the Chilean Government
through the Millennium Science Initiative and the Centers of
Excellence Base Financing Program of CONICYT. CECS is also supported
by a group of private companies which at present includes
Antofagasta Minerals, Arauco, Empresas CMPC, Indura, Naviera
Ultragas and Telef\'{o}nica del Sur. CIN is funded by CONICYT and
Gobierno Regional de Los R\'{\i}os.
\\

\appendix
\section{Conventions}

We summarize here basic formulas mentioned in the text and
conventions~\cite{Manes-85} we use: Let $s_p$ be a face of a simplex
$S$ of dimension $p$ and let $M_q$ be a submanifold of $M$ of
dimension $q$.
\begin{align*}
s_p\times M_q &= (-1)^{pq}M_q\times s_p,\\
 \partial(s_p\times M_q) &= \partial s_p\times M_q
 + (-1)^p s_p\times \partial M_q,
\end{align*}
Let $\alpha$ be a differential form of degree $(p,q)$ in $(dt,dx)$.
\begin{align*}
 \alpha \equiv \alpha_{[p,q]} d^pt \wedge d^qx &=
 (-1)^{pq} \alpha_{[p,q]} d^qx \wedge d^pt\, .\\
\end{align*}
Integration of $\alpha$ is defined by:
\begin{equation*}
 \int_{s_p \times M_q} \alpha = \int_{M_q} \left( \int_{s_p} \alpha_{[p,q]}
 d^p t \right) d^q x.
\end{equation*}
With these conventions Stokes' theorem is as expected:
\begin{gather}
 \int_{s_p \times M_q} d\alpha = (-1)^p \int_{s_p\times \partial M_q}\alpha,\nonumber
 \\
 \int_{s_p \times M_q} \delta\alpha = \int_{\partial s_p\times M_q}\alpha,
 \\  \Rightarrow
 \int_{s_p \times M_q} (d+\delta)\alpha =
 \int_{\partial(s_p \times M_q)} \alpha.
\end{gather}

\bibliography{Lovelock}
\bibliographystyle{unsrt}

\end{document}